\documentclass[12pt]{article}
\usepackage{color}
\usepackage{a4}
\usepackage{latexsym}
\usepackage{cite}
\usepackage{amsfonts}
\usepackage{amsmath}

\usepackage{breqn}
\allowdisplaybreaks

\usepackage{epsfig}
\usepackage{graphicx}

\usepackage{pslatex}
\usepackage[latin1]{inputenc}
\usepackage[T1]{fontenc}

\renewcommand{\theequation}{\thesection.\arabic{equation}}

\def\slash#1{\rlap{\hbox{$\mskip 1 mu /$}}#1}      

\newcommand{\beq}{\begin{equation}}
\newcommand{\eeq}{\end{equation}}
\newcommand{\bea}{\begin{eqnarray}}
\newcommand{\eea}{\end{eqnarray}}

\newcommand{\MSb}{$\overline{\mbox{MS}}$}

\def\as(#1){{\alpha_{\rm s}^{\,#1}}}
\def\ar(#1){{a_{\rm s}^{\,#1}}}

\def\B(#1,#2){{\beta_{#1}^{\,#2}}}

\def\nc{{n_c}}

\def\ca{{C^{}_A}}

\def\cf{{C^{}_F}}

\def\nf{{n^{}_{\! f}}}

\def\S(#1,#2){{{S}_{#1}(#2)}}

\def\gNknlo{\gamma_{N,k}^{\mathcal{D}, (0)}}
\def\gNknnlo{\gamma_{N,k}^{\mathcal{D}, (1)}}
\def\gNknnnlo{\gamma_{N,k}^{\mathcal{D}, (2)}}
\def\gNknnnnlo{\gamma_{N,k}^{\mathcal{D}, (3)}}
\def\gNknnnnnlo{\gamma_{N,k}^{\mathcal{D}, (4)}}

\begin{document}
\setlength{\parskip}{0.2cm}
\setlength{\baselineskip}{0.55cm}

\begin{titlepage}
\noindent
DESY 21-098 \hfill June 2021\\
\vspace{0.6cm}
\begin{center}
{\LARGE \bf 
    Renormalization of non-singlet quark operator \\[1ex] matrix elements for off-forward hard scattering}\\ 
\vspace{1.4cm}
\large
S. Moch$^{\, a}$ and S.~Van Thurenhout$^{\, a}$\\
\vspace{1.4cm}
\normalsize
{\it $^a$II.~Institute for Theoretical Physics, Hamburg University\\
\vspace{0.1cm}
D-22761 Hamburg, Germany}\\
\vspace{1.4cm}
{\large \bf Abstract}
\vspace{-0.2cm}
\end{center}
We calculate non-singlet quark operator matrix elements of deep-inelastic
  scattering in the chiral limit including operators with total derivatives.
  This extends previous calculations with zero-momentum transfer through the
  operator vertex which provides the well-known anomalous dimensions for the
  evolution of parton distributions, as well as calculations in off-forward kinematics utilizing conformal symmetry. Non-vanishing momentum-flow through the operator vertex leads to mixing
  with total derivative operators under renormalization.
  In the limit of a large number of quark flavors $n_f$ and for low moments in full QCD, we determine the anomalous dimension matrix to fifth order 
in the perturbative expansion in the strong coupling $\alpha_s$ 
  in the \MSb-scheme. 
  We exploit consistency relations for the anomalous dimension matrix which follow from the renormalization
  structure of the operators, combined with a direct calculation of the relevant diagrams up to fourth order.
\vspace*{0.3cm}
\end{titlepage}
%
%
\section{Introduction}
\label{sec:intro}

Within the gauge theory of the strong interaction, quantum chromodynamics (QCD), 
important nonperturbative information about the hadron structure is obtained from matrix
elements of local operators between states with the same or different momenta. 
Depending on the momentum transfer, such operator matrix elements (OMEs) 
are used to describe the parton distribution functions (PDFs) in forward kinematics, 
their off-forward counter-parts, the generalized parton distributions (GPDs) 
or distribution amplitudes (DAs) for vacuum-to-hadron transitions, such as the pion form factor. Extensive reviews on the description of hadron structure using GPDs and DAs can be found in e.g. \cite{Diehl:2003ny} and \cite{Belitsky:2005qn}.
PDFs or GPDs can be extracted from data for hard inclusive and semi-inclusive reactions with identified particles in the
final state, collected for example in deep-inelastic scattering (DIS) in the past at the HERA collider~\cite{Abramowicz:2015mha,Accardi:2016ndt} 
or in the future through the research program at the planned Electron Ion Collider (EIC)~\cite{Boer:2011fh,AbdulKhalek:2021gbh}.
Various DAs can be obtained from data taken in high-intensity, medium energy experiments, for instance at Belle II~\cite{Kou:2018nap}.
All such nonperturbative quantities are also accessible from first principles in QCD simulations of
the hadron structure on the lattice, see e.g.~\cite{Gockeler:2004wp, Gockeler:2010yr} 
and for recent progress~\cite{Braun:2015axa,Braun:2016wnx,Bali:2018zgl,Bali:2019dqc,Harris:2019bih,Alexandrou:2020sml}.

The scale-dependence of such distributions is governed by the renormalization
group equations for the corresponding operators and can be computed order by order in the strong coupling $\alpha_s$ in perturbative QCD.
We focus here on the non-singlet quark OMEs relevant in DIS in the chiral
limit and include mixing with operators involving total derivatives.
In a given basis of local operators this implies renormalization with a triangular mixing matrix
where the diagonal entries are the forward anomalous dimensions, well-known
from inclusive DIS, but the off-diagonal elements of the renormalization
matrix require a separate calculation.

The anomalous dimensions of non-singlet quark operators as functions of the Mellin moment $N$, 
which coincides with the Lorentz spin of the operator, are completely known up to three loops~\cite{Gross:1973ju, Floratos:1977au, Moch:2004pa},
and, likewise the corresponding splitting functions in $x$-space. 
At the four-loop level, low fixed moments~\cite{Velizhanin:2011es,Velizhanin:2014fua,Ruijl:2016pkm} and 
complete results in the limit of a large number $n_f$ of quark flavors~\cite{Gracey:1994nn,Davies:2016jie} 
have been obtained. 
More recently, fixed moments up to $N=16$ have been computed and 
used to determine complete analytic results for a general $SU(n_c)$ gauge theory in the planar limit, 
i.e. in the limit of a large $n_c$~\cite{Moch:2017uml}.
Partial information consisting of the large $n_f$ limit~\cite{Gracey:1994nn} and of low moments~\cite{Herzog:2018kwj} is available even at five loops.

In off-forward kinematics, the three-loop evolution kernel for flavor-non-singlet operators is known as well~\cite{Braun:2017cih}. 
The computation has exploited conformal symmetry~\cite{Braun:2003rp} of the QCD Lagrangian which
was first utilized in pioneering work for the two-loop radiative corrections~\cite{Mueller:1993hg,Belitsky:1998gc}. 
In $d=4-2\epsilon$ dimensions and adopting the modified minimal subtraction (\MSb) scheme 
conformal symmetry in QCD is exact for large $n_f$ at the critical coupling.
In the physical four-dimensional theory the renormalization group equations
then inherit a conformal symmetry so that the generators of the conformal
transformations and the evolution kernel commute~\cite{Braun:2013tva}.
Consistency relations following from the conformal algebra allow to restore the
$l$-loop triangular mixing matrix for the off-forward anomalous dimensions 
from the $l$-loop forward anomalous dimensions and an $(l-1)$-loop
result for a so-called conformal  anomaly~\cite{Mueller:1991gd,Braun:2016qlg,Braun:2017cih}. 
In moment space, the off-forward anomalous dimension depends on the Lorentz
spin $N$ as well as on the number $k$ of total derivatives acting on the operator. 
In momentum space, the corresponding conjugate variables are the momentum fraction $x$ 
carried by the struck parton and an additional kinematical variable such as
the skewedness parameter of the process.

Fixed moments of non-singlet quark OMEs at three-loop order have also been computed
in the \MSb\ renormalization scheme as well as in alternative ones, 
such as the regularization invariant (RI) scheme, 
which are suitable for a direct application to available lattice results~\cite{Gracey:2009da,Kniehl:2020nhw}.
The choice for a basis of the renormalized local operators in these works 
is different from the one in the conformal approach~\cite{Mueller:1993hg},
so that a comparison between the fixed moment results of~\cite{Gracey:2009da,Kniehl:2020nhw} 
and those of~\cite{Braun:2017cih,Mueller:1993hg,Belitsky:1998gc} 
is not immediate, requiring additional computational steps.

In the present article, we review the renormalization of non-singlet quark
operators with particular emphasis on possible choices for their bases. 
In this way, we connect to different results that have appeared as of yet unrelated 
in the literature.
We perform explicit computations of the relevant OMEs up to four loops for
non-zero momentum transfer through the operator vertex 
and derive a number of consistency relations for the respective anomalous 
dimensions which govern the mixing associated with total derivative operators.
This checks and extends previous calculations for the leading-$n_f$ terms 
in the evolution of flavor non-singlet operators in off-forward kinematics 
up to five loops.

The article is organized as follows:
In Sec.~\ref{sec:frame} we set the stage, review the different basis choices
for spin-$N$ local non-singlet quark operators used in the literature and
discuss their renormalization together with particular properties of the
anomalous dimensions. 
Details of the computation are given in Sec.~\ref{sec:calc} and results
for the moments of the complete mixing matrix up to five loops for the leading-$n_f$ terms are presented in Sec.~\ref{sec:Nres}. In Sec.~\ref{sec:beyondNF}, we list some results beyond the leading-$\nf$ limit, including the second order mixing matrix in the planar limit and fixed numerical moments in full QCD up to fifth order. The corresponding moments for a general gauge group $SU(\nc)$ are deferred to App.~\ref{sec:appA}.
Conclusions and an outlook are given in Sec.~\ref{sec:conclusion}.

%
\section{Theoretical framework}
\label{sec:frame}
\setcounter{equation}{0}

We consider the renormalization of spin-$N$ local non-singlet quark operators
\begin{equation}
\label{eq:OpDef}
    \mathcal{O}^{NS}_{\mu_1 \dots \mu_{N}} \,=\, 
    \mathcal{S}\, \overline{\psi}\lambda^{\alpha}\gamma_{\mu_1} D_{\mu_2} \dots D_{\mu_{N}}\psi\, ,
\end{equation}
where $\psi$ represents the quark field, $D_{\mu} = \partial_{\mu} - i g_s A_{\mu}$
the standard QCD covariant derivative, and $\lambda^{\alpha}$ the generators
of the flavor group $SU(n_f)$. 
Since we focus on the leading-twist contributions, 
we symmetrize the Lorentz indices and take the traceless part, indicated by ${\mathcal S}$. 
{This projects the twist-two contribution, see e.g.~\cite{Blumlein:1999sc}.}

The anomalous dimensions of interest are determined by considering  
spin-averaged OMEs obtained by inserting the respective operators in two-point functions, 
\begin{equation}
\label{eq:generalOME}
    \langle \psi(p_1) | \mathcal{O}_{\mu_1 \dots \mu_N}^{NS}(p_3) | \overline{\psi}(p_2)\rangle 
    \,,
\end{equation}
with quarks and anti-quarks of momenta $p_1$ and $p_2$ as external fields and all momenta are incoming, $\sum_{i=1}^3 p_i = 0$.
Choosing $p_3=0$ leads to zero momentum-flow through the
operator and the OMEs are renormalized with the standard forward anomalous
dimensions, see e.g.~\cite{Moch:2017uml}. 
For the general case $p_3 \neq 0$ there is a momentum transfer through the operator vertex, 
which implies mixing between the operators in Eq.~(\ref{eq:OpDef}) and 
additional total derivative operators. 

Next we will introduce {three} 
sets of bases commonly used in the literature.
One is built from an expansion in Gegenbauer polynomials, which is used in the 
conformal approach, and the other {two 
are based on counting powers of (total) derivatives}.

\subsection{The Gegenbauer basis}
We start our discussion by introducing the renormalized non-local light-ray operators $[\mathcal{O}]$. 
These act as generating functions for local operators, see \cite{Braun:2017cih}, as 
\begin{equation}
\label{eq:lightcone}
    [\mathcal{O}](x;z_1,z_2) \,=\, 
    \sum\limits_{m,k} \frac{z_1^m z_2^k}{m!\, k!}
    \left[\overline{\psi}(x) (\stackrel{\leftarrow}{D} \cdot n)^m \slash{n} (n \cdot \stackrel{\rightarrow}{D})^k \psi(x)\right]
    \, .
\end{equation}
Here, $n$ is an arbitrary light-like vector. 
For simplicity, the $x$-dependence will be omitted in the following, 
writing $\mathcal{O}(z_1,z_2) \equiv \mathcal{O}(0;z_1,z_2)$. 
The renormalization group equation for these light-ray operators can be written as 
\begin{equation}
    \Big(\mu^2 \partial_{\mu^2} + \beta(a_s) \partial_{a_s} + \mathcal{H}(a_s)\Big)[\mathcal{O}](z_1,z_2) = 0
    \, ,
\end{equation}
with $\mu$ the renormalization scale, $a_s = \alpha_s/(4\pi)$ the strong coupling 
and $\beta(a_s)$ the QCD beta-function. 
The evolution operator $\mathcal{H}(a_s)$ is an integral operator and acts on the
light-cone coordinates of the fields~\cite{Balitsky:1987bk} 
\begin{equation}
    \mathcal{H}(a_s)[\mathcal{O}](z_1,z_2) \,=\, \int_{0}^{1}d\alpha \int_{0}^{1}d\beta \: h(\alpha,\beta) [\mathcal{O}](z_{12}^{\alpha},z_{21}^{\beta})
\end{equation}
with $z_{12}^{\alpha} \equiv z_1(1-\alpha) + z_2\alpha$ and the evolution kernel $h(\alpha,\beta)$. 
The moments of the evolution kernel correspond to the anomalous dimensions of the local operators in Eq.~(\ref{eq:lightcone}) 
\begin{equation}
    \label{eq:gammaN}
    \gamma_{N,N} = \int_{0}^{1}d\alpha \int_{0}^{1}d\beta\, (1-\alpha-\beta)^{N-1} h(\alpha,\beta)
    \, ,
\end{equation}
where $N$ represents the total number of covariant derivatives appearing 
in the operator, i.e. $N=m+k$ in Eq.~(\ref{eq:lightcone}). The forward anomalous dimensions will be represented by $\gamma_{N,N}$, and there is a shift $N \rightarrow N+1$ compared to the literature which is related to the operator definition.

The light-ray operators in Eq.~(\ref{eq:lightcone}) admit an expansion in a
basis of local operators in terms of Gegenbauer polynomials, see e.g. \cite{Efremov:1978rn,Belitsky:1998gc},
\begin{equation}
\label{eq:Gbasis}
    \mathcal{O}_{N,k}^{\mathcal{G}} = (\partial_{z_1}+\partial_{z_2})^k
    C^{3/2}_N
    \Big(\frac{\partial_{z_1}-\partial_{z_2}}{\partial_{z_1}+\partial_{z_2}}\Big)\mathcal{O}(z_1,z_2)\biggr\vert_{z_1=z_2=0}
    \, , 
\end{equation}
where $k \geq N$ is the total number of derivatives and 
we use the superscript $\mathcal{G}$ to denote the operators in the Gegenbauer basis. 
For any given $N$, the operator of lowest dimension, $\mathcal{O}_{N,N}^{\mathcal{G}}$, is a conformal operator.
The Gegenbauer polynomials can be written in terms of the hypergeometric function $_2F_1$ as~\cite{olver10}
\begin{eqnarray}
    C_N^{\nu}(z) &=& \frac{(2\nu)_N}{N!}{_2}F_1\Big(-N,N+2\nu;\nu+\frac{1}{2};\frac{1}{2}-\frac{z}{2}\Big) 
    \nonumber\\
    &=& 
    \frac{\Gamma(\nu+1/2)}{\Gamma(2\nu)}\, \sum\limits_{l=0}^{N}\,
    (-1)^l\frac{\Gamma(2\nu+N+l)}{l!\, (N-l)! \Gamma(\nu+1/2+l)}\, 
    \Big(\frac{1}{2}-\frac{z}{2}\Big)^l
    \, ,
\end{eqnarray}
where the series representation in terms of the Gamma function will be more convenient for our purposes.
The Gegenbauer polynomial with the differential operators then gives
\begin{eqnarray}
  \label{eq:Gexp}
    (\partial_{z_1}+\partial_{z_2})^k
    C_{N}^{3/2}\Big(\frac{\partial_{z_1}-\partial_{z_2}}{\partial_{z_1}+\partial_{z_2}}\Big) 
    \,=\, 
    \frac{1}{2N!}\sum\limits_{l=0}^{N}(-1)^l \binom{N}{l}\frac{(N+l+2)!}{(l+1)!}\sum\limits_{j=0}^{k-l}\binom{k-l}{j}(\partial_{z_1})^{k-l-j}(\partial_{z_2})^{l+j}
    \, ,
    \nonumber\\
\end{eqnarray}
with $k \geq N$. 
The renormalized operators $[\mathcal{O}_{N,k}^{\mathcal{G}}]$ obey the evolution equation
\begin{equation}
\label{eq:evolG}
\Big(\mu^2 \partial_{\mu^2} +
\beta(a_s)\partial_{a_s}\Big)[\mathcal{O}_{N,k}^{\mathcal{G}}] \,=\,
\sum\limits_{j=0}^{N}\gamma_{N,j}^{\mathcal{G}}[\mathcal{O}_{j,k}^{\mathcal{G}}]
\, ,
\end{equation}
where the mixing of the operators is manifest. 
The anomalous dimension matrix~\footnote{We represent the elements of the mixing matrix for spin-($N+1$) operators as $\gamma_{N,k}$ and the matrix itself as $\hat{\gamma}_{N+1}$.}, denoted by  $\hat{\gamma}_{N}^{\mathcal{G}}$, is triangular, 
i.e. its elements $\gamma_{N,j}^{\mathcal{G}} = 0$ if $j>N$.
It can be computed in QCD perturbation theory and its diagonal elements correspond  
to the standard forward anomalous dimensions $\gamma_{N,N}$~\cite{Moch:2017uml}. Furthermore, the superscript ${\mathcal{G}}$ can be dropped for $\gamma_{N,N}$ since they do not depend on the particular choice for the basis of additional total derivatives.
The matrix $\hat{\gamma}_{N}^{\mathcal{G}}$ is currently known up to the three-loop level~\cite{Braun:2017cih} 
and we will discuss these results in more detail in Sec.~\ref{sec:Nres}.

\subsection{The total derivative basis}
Another basis for the quark operators, which directly generalizes Eq.~(\ref{eq:OpDef}) is
\begin{equation}
\label{eq:derivativeBasis}
    \mathcal{O}_{p,q,r}^{\mathcal{D}} \,=\, 
    \mathcal{S}\, \partial^{\mu_1}\dots \partial^{\mu_p}\, ((D^{\nu_1}\dots D^{\nu_q}\overline{\psi})\, \lambda^{\alpha}\gamma_{\mu}\, 
    (D^{\sigma_1}\dots  D^{\sigma_r}\psi))
    \, ,
\end{equation}
see e.g.~\cite{Gracey:2011zn,Gracey:2011zg} 
{and \cite{Geyer:1982fk,Blumlein:1999sc}}.
Here the superscript $\mathcal{D}$ indicates that the operators are written in the 
total derivative basis and the indices $p$, $q$ and $r$ count the powers of the respective derivatives.
In practical calculations, all Lorentz indices are contracted with a light-like vector $\Delta$, such that the indices $p$, $q$ and $r$ in Eq.~(\ref{eq:derivativeBasis}) identify a given operator uniquely. 
Because of the chiral limit, the partial derivatives act as
\begin{equation}
\label{eq:partialAct}
    \mathcal{O}_{p,q,r}^{\mathcal{D}} \,=\, \mathcal{O}_{p-1,q+1,r}^{\mathcal{D}} + \mathcal{O}_{p-1,q,r+1}^{\mathcal{D}}
    \, .
\end{equation}
For the bare operators this defines a recursion, which is solved by
\begin{equation}
\label{eq:operatorRec}
    \mathcal{O}_{p,q,r}^{\mathcal{D}} \,=\, 
    \sum\limits_{i=0}^p\, \binom{p}{i}\, \mathcal{O}_{0,p+q-i,r+i}^{\mathcal{D}}\, .
\end{equation}
Another consequence of the chiral limit is that left and right derivative
operators renormalize with the same renormalization constants
\begin{eqnarray}
  \label{eq:renormPattern}
  \mathcal{O}_{p,0,r}^{\mathcal{D}} &=& \sum\limits_{j=0}^{r}\, Z_{r,r-j}\, [\mathcal{O}_{p+j,0,r-j}^{\mathcal{D}}]
  \, , \\
  \mathcal{O}_{p,q,0}^{\mathcal{D}} &=& \sum\limits_{j=0}^{q}\, Z_{q,q-j}\, [\mathcal{O}_{p+j,q-j,0}^{\mathcal{D}}]
  \, .
\end{eqnarray}
The anomalous dimensions $\gamma_{N,k}^{\mathcal{D}}$ governing the scale dependence
of these operators are derived from the $Z$-factors by
\begin{equation}
\label{eq:gamZ}
\gamma_{N,k}^{\mathcal{D}} \,=\,
-\,\bigg( \,\frac{d }{d\ln\mu^2 }\; Z_{N,j} \bigg)\, Z_{j,k}^{\,-1}
\, .
\end{equation}
The mixing matrix is also triangular in the total derivative basis ($\gamma_{N,k}^{\mathcal{D}} = 0$ if $k > N$)
and, as was the case for the Gegenbauer basis, the diagonal elements are the standard forward anomalous dimensions $\gamma_{N,N}$~\cite{Moch:2017uml}, where we have also dropped the superscript ${\mathcal{D}}$ due to the basis independence.

It is possible to relate the two operator bases using the light-ray operators in Eq.~(\ref{eq:lightcone}) as generating functions.
Starting from the bare operators we have
\begin{equation}
    \mathcal{O}(z_1,z_2) \,=\, \sum\limits_{m,k}\, \frac{z_1^m z_2^k}{m!\, k!}\, \mathcal{O}_{0,m,k}^{\mathcal{D}}
    \, ,
\end{equation}
so that Eqs.~(\ref{eq:Gbasis}), (\ref{eq:Gexp}) and (\ref{eq:operatorRec}) give
\begin{equation}
\label{eq:operatorG}
    \mathcal{O}_{N,k}^{\mathcal{G}} \,=\, \frac{1}{2N!}\,\sum\limits_{l=0}^{N}\,(-1)^l\binom{N}{l}\,\frac{(N+l+2)!}{(l+1)!}\,\mathcal{O}_{k-l,0,l}^{\mathcal{D}}
    \, .
\end{equation}
The evolution equation for the operators $\mathcal{O}_{N,k}^{\mathcal{G}}$ in Eq.~(\ref{eq:evolG}) 
relates the anomalous dimension matrices in the two bases
\begin{equation}
    \sum\limits_{j=0}^N\, \gamma_{N,j}^{\mathcal{G}}\, [\mathcal{O}_{j,N}^{\mathcal{G}}]
    \,=\, 
    \frac{1}{2N!}\, \sum\limits_{l=0}^{N}\, (-1)^l\binom{N}{l}\frac{(N+l+2)!}{(l+1)!}\, 
    \sum\limits_{j=0}^{l}\, \gamma_{l,l-j}^{\mathcal{D}}\, [\mathcal{O}_{N-l+j,0,l-j}^{\mathcal{D}}]
    \, .
\end{equation}
Using Eq.~(\ref{eq:operatorG}) for renormalized operators the left-hand side can be expanded
in the operators of the total derivative basis,
\begin{eqnarray}
{\lefteqn{
    \sum\limits_{j=0}^{N}\, \gamma_{N,j}^{\mathcal{G}}\, 
    \Bigg(\frac{1}{2j!}\sum\limits_{l=0}^{j}\,(-1)^l \binom{j}{l}\,
    \frac{(j+l+2)!}{(l+1)!}\, [\mathcal{O}_{N-l,0,l}^{\mathcal{D}}]\Bigg) 
    \,\, }}
\nonumber\\ &=&
    \frac{1}{2N!}\, \sum\limits_{j=0}^{N}\, (-1)^j\binom{N}{j}\,
    \frac{(N+j+2)!}{(j+1)!}\,
    \sum\limits_{l=0}^{j}\gamma_{j,l}^{\mathcal{D}}\, [\mathcal{O}_{N-l,0,l}^{\mathcal{D}}]
    \, ,
\end{eqnarray}
and upon comparing the coefficients of the operators $[\mathcal{O}_{N-l,0,l}^{\mathcal{D}}]$, we find
\begin{equation}
\label{eq:CFvsum}
    \sum\limits_{j=0}^{N}\, (-1)^j\frac{(j+2)!}{j!}\, \gamma_{N,j}^{\mathcal{G}} \,=\,
    \frac{1}{N!}\, \sum\limits_{j=0}^{N}\,
    (-1)^j\binom{N}{j}\frac{(N+j+2)!}{(j+1)!}\, \sum\limits_{l=0}^{j}\, \gamma_{j,l}^{\mathcal{D}}
    \, .
\end{equation}
To get the expression on the left-hand side, we have evaluated the sum
\begin{equation}
    \sum\limits_{l=0}^{j}\, (-1)^{l}\binom{j}{l}\, \frac{(j+l+2)!}{(l+1)!} \,=\,
    (-1)^j(j+2)!
    \,.
\end{equation}

{
\subsection{The Geyer basis}
{The local operators introduced by B.~Geyer in  \cite{Geyer:1982fk,Blumlein:1999sc} are expressed in a basis different from the one in Eq.~(\ref{eq:renormPattern}), 
but the quoted anomalous dimensions can be related to  $\gamma_{N,k}^{\mathcal{D}}$ in Eq.~(\ref{eq:gamZ}).
}
The basis for the local operators used in those references is
\begin{equation}
\label{eq:geyerBasis}
    \mathcal{O}_{N,k}^{\mathcal{B}} \,=\, \overline{\psi}\gamma(\stackrel{\rightarrow}{D}+\stackrel{\leftarrow}{D})^{N-k}(\stackrel{\leftarrow}{D}-\stackrel{\rightarrow}{D})^k\psi
    \, .
\end{equation}
with $N,k$ odd. The superscript $\mathcal{B}$ indicates that the operators are written in the Geyer basis. The contraction with an arbitrary light-like vector is understood, i.e.
\begin{equation}
    \gamma \,\equiv\, \Delta^{\mu}\:\gamma_{\mu}\, , \qquad
    D \,\equiv\, \Delta^{\mu}\:D_{\mu}\, ,
\end{equation}
and $\Delta^2=0$.

We now want to relate the operators, and the corresponding anomalous dimensions, in the $\mathcal{B}$-basis to those in the derivative basis. For this we use
\begin{equation}
\label{eq:ONk}
    \mathcal{O}^{\mathcal{D}}_{0,N-k,k} = (-1)^k\sum_{j=0}^{k}\binom{k}{j}\mathcal{O}^{\mathcal{D}}_{j,N-j,0},
\end{equation}
which can be derived in the same way as Eq.~(\ref{eq:operatorRec}). It is then straightforward to see that, for $k=0$, the operators defined above correspond to total derivatives of the vector current, i.e.
\begin{equation}
    \mathcal{O}^{\mathcal{B}}_{N,0} = \mathcal{O}^{\mathcal{D}}_{N,0,0}.
\end{equation}
For arbitrary $k$-values, we can use the binomial theorem (twice) together with Eq.~(\ref{eq:ONk}). This leads to the following relation between the bare operators
\begin{equation}
    \mathcal{O}^{\mathcal{B}}_{N,k} = \sum_{i=0}^{N-k}\sum_{j=0}^{k}(-1)^{i}\binom{N-k}{i}\binom{k}{j}\sum_{l=0}^{i+j}(-1)^l\binom{i+j}{l}\mathcal{O}^{\mathcal{D}}_{l,N-l,0}.
\end{equation}
The corresponding relation for the renormalized operators reads
\begin{align}
\label{eq:geyerRen}
    [\mathcal{O}^{\mathcal{B}}_{N,k}] =& \sum_{i=0}^{N-k}\sum_{j=0}^{k}(-1)^i\binom{N-k}{i}\binom{k}{j}\sum_{l=0}^{i+j}(-1)^l\binom{i+j}{l}\sum_{m=0}^{N-l}(-1)^m\:\gamma_{N-l,m}^{\mathcal{D}}\nonumber\\&\times\sum_{n=0}^m(-1)^n\binom{m}{n}[\mathcal{O}^{\mathcal{D}}_{N-m+n,0,m-n}].
\end{align}
Like in the Gegenbauer basis, we now focus on the diagonal operators $\mathcal{O}^{\mathcal{B}}_{N,N}$, the evolution equation for which is
\begin{equation}
    \mu^2 \frac{d}{d\mu^2}\mathcal{O}_{N,N}^{\mathcal{B}} = \sum_{k=0}^{N}\frac{1-(-1)^k}{2}\gamma_{N,k}^{\mathcal{B}}[\mathcal{O}^{\mathcal{B}}_{k,N}]
\end{equation}
with $N$ odd. Like in the other operator bases, 
the anomalous dimension matrix $\hat{\gamma}_{N}^{\mathcal{B}}$ 
is triangular ($\gamma_{N,k}^{\mathcal{B}} = 0$ if $k > N$) 
and in analogy to Eq.~(\ref{eq:CFvsum}) $\gamma_{N,k}^{\mathcal{B}}$ can be 
related to the mixing matrices in the Gegenbauer and the total derivative bases. 
Details will be given below in Sec.~\ref{1loopRES}.
}

\subsection{Constraints on the anomalous dimensions}
\label{sec:constraintsADM}
The elements of the mixing matrices for the operators in the total derivative basis 
are not all independent, but subject to particular constraints, 
which define useful relationships between them in the chiral limit.
Starting from Eq.~(\ref{eq:partialAct}) we can derive the following relation 
by acting $N$ times with a partial derivative on the bare operator  $\mathcal{O}_{N,0,0}^{\mathcal{D}}$
\begin{equation}
\label{bareREL}
    \mathcal{O}_{0,N,0}^{\mathcal{D}}-(-1)^N \sum_{j=0}^{N}(-1)^j\binom{N}{j}\mathcal{O}_{j,0,N-j}^{\mathcal{D}} \,=\, 0
    \, .
\end{equation}
Upon using the renormalization equations (\ref{eq:renormPattern}), this leads to a connection between the renormalized operators
\begin{equation}
    \sum_{k=0}^{N}\sum_{j=k}^{N}\Bigg\{(-1)^k\binom{j}{k}Z_{N,j}-(-1)^j\binom{N}{j}Z_{j,k}\Bigg\}[\mathcal{O}_{N-k,0,k}^{\mathcal{D}}] \,=\, 0
    \, .
\end{equation}
Now, since the coefficient of $[\mathcal{O}_{N-k,0,k}^{\mathcal{D}}]$ has to vanish 
for each value of $k$, this becomes a relation between the renormalization constants
\begin{equation}
    \forall k:\qquad 
    \sum_{j=k}^{N}\Bigg\{(-1)^k\binom{j}{k}Z_{N,j}-(-1)^j\binom{N}{j}Z_{j,k}\Bigg\} \,=\, 0
    \, ,
\end{equation}
which is equivalent to a relation between sums of elements of the mixing matrix
$\hat{\gamma}_{N}^{\mathcal{D}}$,
\begin{equation}
\label{mainGamma}
    \forall k: \qquad 
    \sum_{j=k}^{N}\Bigg\{(-1)^k\binom{j}{k}\gamma_{N,j}^{\mathcal{D}}-(-1)^j\binom{N}{j}\gamma_{j,k}^{\mathcal{D}}\Bigg\} \,=\, 0
    \, .
\end{equation}
This is a general statement, valid to all orders in $a_s$. 
It can be considered as a consequence of parity conservation. If one takes a basis with $\partial_{z_1}-\partial_{z_2}$ as in Eq.~(\ref{eq:Gbasis}), i.e., "left minus right (l-r)" derivatives, each such derivative changes parity so that the operators with an even number of "l-r" derivatives do not mix with operators with an odd number of "l-r" derivatives.~\footnote{We thank V.~Braun for a discussion on this aspect.}

By putting $k=N-1$ we can relate the next-to-diagonal elements of the mixing matrix to the forward anomalous dimensions $\gamma_{N,N}$, cf.~Eq.~(\ref{eq:gammaN})~\footnote{The low-$N$ version of this relation was pointed out to
us by J. Gracey in a private communication.},
\begin{equation}
    \label{NTD}
    \gamma_{N,N-1}^{\mathcal{D}} \,=\, \frac{N}{2}\left(\gamma_{N-1,N-1}-\gamma_{N,N}\right)
    \, ,
\end{equation}
where, as above, the superscript ${\mathcal{D}}$ can be omitted for $\gamma_{N,N}$.

The case $k=0$ in Eq.~(\ref{mainGamma}) gives
\begin{equation}
\label{mainK0}
    \sum_{j=0}^{N}\Bigg\{\gamma_{N,j}^{\mathcal{D}}-(-1)^j\binom{N}{j}\gamma_{j,0}^{\mathcal{D}}\Bigg\} \,=\, 0
    \, ,
\end{equation}
which relates the sum of the elements in the $N$-th row of the mixing matrix to the conjugate $\mathcal{C}\, \gamma_{N,0}^{\mathcal{D}}$ of the last column defined as 
\begin{equation}
\label{g0}
    \mathcal{C}\, \gamma_{N,0}^{\mathcal{D}} \,\equiv\, \sum_{i=0}^{N} (-1)^i\binom{N}{i}\, \gamma_{i,0}^{\mathcal{D}}
    \, ,
\end{equation}
see e.g. \cite{Vermaseren:1998uu}.

With the help of Eq.~(\ref{mainK0}) for the last column of the mixing matrix 
another relation between the anomalous dimensions in the total derivative basis and the Gegenbauer one can be obtained. 
Substituting Eq.~(\ref{mainK0}) into Eq.~(\ref{eq:CFvsum}) one finds
\begin{equation}
\label{eq:GtoD}
    \sum_{j=0}^{N}(-1)^j\frac{(j+2)!}{j!}\gamma_{N,j}^{\mathcal{G}} \,=\, \frac{1}{N!}\sum_{j=0}^{N}(-1)^j\binom{N}{j}\frac{(N+j+2)!}{(j+1)!}\sum_{l=0}^{j}(-1)^l\binom{j}{l}\gamma_{l,0}^{\mathcal{D}}
    \, .
\end{equation}

Considering now arbitrary $k$ in Eq.~(\ref{mainGamma}) results in
\begin{equation}
\label{mainConj}
    \gamma_{N,k}^{\mathcal{D}} \,=\, 
    \binom{N}{k}\sum_{j=0}^{N-k}(-1)^j \binom{N-k}{j}\gamma_{j+k,j+k} 
    + \sum_{j=k}^N (-1)^k \binom{j}{k} \sum_{l=j+1}^N (-1)^l \binom{N}{l} \gamma_{l,j}^{\mathcal{D}}
    \, .
\end{equation}
This is a useful relation for several reasons. 
Its derivation only relies on the chiral limit, which imposes constraints on the renormalization structure of the operators. It can, therefore, be used as an order-independent consistency check, which any expression for $\gamma_{N,k}^{\mathcal{D}}$ has to obey. 
Alternatively, Eq.~(\ref{mainConj}) provides a path for the construction of the full mixing matrix $\hat{\gamma}_{N}^{\mathcal{D}}$ 
from the knowledge of the forward anomalous dimensions $\gamma_{N,N}$ 
and the last column $\gamma_{N,0}^{\mathcal{D}}$ at any order of perturbation theory.

Thus, with partial information being available even to five-loop order, one can construct an ansatz for the off-diagonal elements 
and use Eq.~(\ref{mainConj}) to test its self-consistency. 
In this case, $\gamma_{N,0}^{\mathcal{D}}$ serves as boundary condition 
and Eq.~(\ref{NTD}) as a cross-check.
This leads to the following 4-step algorithm for the construction of the mixing matrix:

\begin{enumerate}
    \item Starting from the forward anomalous dimensions $\gamma_{N,N}$ and the bare OMEs in Eq.~(\ref{eq:generalOME}), one determines the all-$N$ expressions for the next-to-diagonal $\gamma_{N,N-1}^{\mathcal{D}}$ and the last column entries $\gamma_{N,0}^{\mathcal{D}}$ of the mixing matrix. In the next section, it will be detailed how to relate the last column entries to the calculation of Feynman diagrams.

    \item Next, one calculates the sum
    \begin{equation}
        \binom{N}{k}\, \sum_{j=0}^{N-k}\, (-1)^j \binom{N-k}{j}\, \gamma_{j+k,j+k}
        \, .
    \end{equation}
    Based on the structure of the result, one can then make an ansatz for the off-diagonal elements.
    \item One calculates the double sum 
    \begin{equation}
    \label{eq:DoubleSum}
        \sum_{j=k}^N\, (-1)^k \binom{j}{k}\, \sum_{l=j+1}^N\, (-1)^l \binom{N}{l}\, \gamma_{l,j}^{\mathcal{D}}
    \end{equation}
    with the chosen ansatz and collects everything into Eq.~(\ref{mainConj}). This leads to a system of equations in the unknown coefficients of the ansatz. It should be mentioned that Eq.~(\ref{mainConj}) by itself is not enough to determine a unique solution for $\gamma_{N,k}^{\mathcal{D}}$. In particular, some structures might be self-conjugate, like e.g.
    \begin{equation}
        \sum_{j=k}^N\, (-1)^k \binom{j}{k}\, \sum_{l=j+1}^N\, (-1)^l \binom{N}{l}\, \frac{1}{l-j} = -\frac{1}{N-k}.
    \end{equation}
    Uniqueness of the solution is saved however by the fact that there is a boundary condition, namely the expression for $\gamma_{N,k}^{\mathcal{D}}$ has to agree with the previously found expression for $\gamma_{N,0}^{\mathcal{D}}$ from step 1.
    \item If the system of equations can be solved consistently, one has determined the final expression for the off-diagonal terms of the mixing matrix $\hat{\gamma}_{N}^{\mathcal{D}}$. If not, the structure of the remaining terms in Eq.~(\ref{mainConj}) can be used to adapt the ansatz, leading one back to step 3.
\end{enumerate}

This approach will be applied and described in further detail below.

%
\section{Calculation in Mellin $N$-space}
\label{sec:calc}
\setcounter{equation}{0}

We will describe the computation of Feynman diagrams for the matrix elements 
of operators in Eq.~(\ref{eq:OpDef}) along with the necessary details for the renormalization.
The constraints for the anomalous dimensions in Eq.~(\ref{mainGamma}) and 
the constructive algorithm to determine the mixing matrix lead to particular types of symbolic sums and 
techniques for their evaluation will be reviewed as well.

\subsection{Calculating Feynman diagrams}
\begin{figure}[t]
\centering
\includegraphics[width=0.5\textwidth]{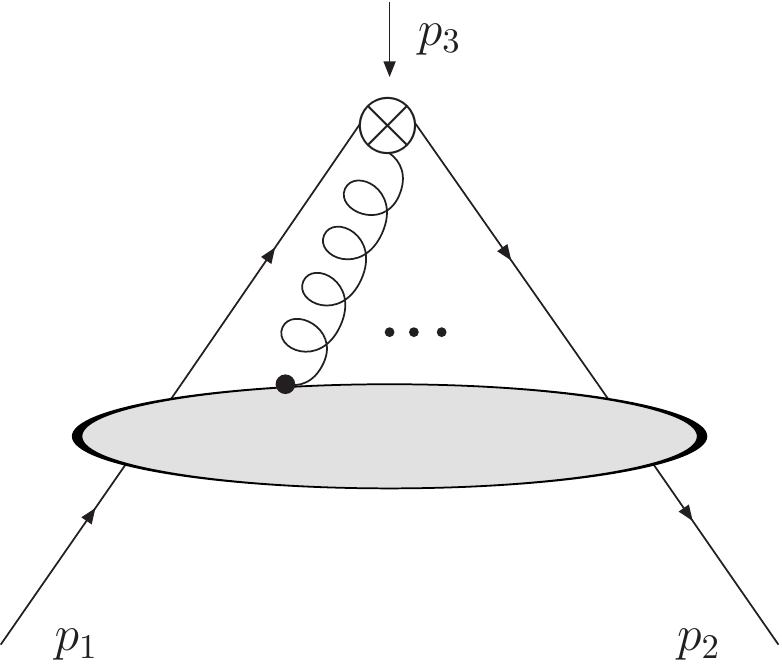}
\caption{\label{figGreenFun}
The Green's function 
$\langle \psi(p_1) | O_{\mu_1\dots \mu_N}^{NS}(p_3) | \overline{\psi}(p_2)\rangle$ 
with momentum $p_3=-p_1-p_2$ flowing through the operator vertex. 
For simplicity, we set $p_2 \equiv 0$.}
\end{figure}

We calculate the spin-averaged OMEs defined in Eq.~(\ref{eq:generalOME}). 
To ensure tracelessness and symmetry of the Lorentz indices, the OMEs are contracted with a tensor of light-like $\Delta$,
\begin{equation}
    \label{eq:theDeltas}
    \Delta^{\mu_1}\dots \Delta^{\mu_N}
    \, ,
\end{equation}
and $\Delta^2=0$. 
OMEs in off-forward kinematics require momentum-flow through the operator vertex, $p_3 \neq 0$, see Fig.~\ref{figGreenFun}.
For simplicity but without loss of generality, one can choose $p_2=0$, 
so that the calculated OMEs are of the form
\begin{equation}
\label{eq:OMEs}
    \Delta^{\mu_1}\dots \Delta^{\mu_N}\, \langle \psi(p_1) | O_{\mu_1\dots \mu_N}^{NS}(-p_1) | \overline{\psi}(0) \rangle
    \, .
\end{equation}
With this momentum configuration the OMEs in Eq.~(\ref{eq:generalOME}), which are a priori three-point functions reduce effectively to two-point ones, i.e. one is left with massless propagator-type Feynman diagrams. These can be efficiently calculated using computer algebra methods as will be detailed next.

The computations follow a well-established workflow. The Feynman
diagrams are generated using {\sc Qgraf} \cite{Nogueira:1991ex}, the output of which 
is then directed to a {\sc Form} \cite{Vermaseren:2000nd,Kuipers:2012rf} program to
determine the topologies and to compute the color factors of the diagrams, the latter part based on the algorithms presented in \cite{vanRitbergen:1998pn}. 
The necessary Feynman rules are given for example in~\cite{Moch:2017uml}.
For computational efficiency, diagrams of the same topology and color factor are grouped together in so-called meta-diagrams, cf.~\cite{Herzog:2016qas} for 
further details on the use of {\sc Form} in Feynman diagram calculations.
The handling of the meta-diagrams is done with the database program {\sc Minos}~\cite{Vermaseren:1997qr}. 
The actual diagram calculations are then performed using the {\sc Forcer} program~\cite{Ruijl:2017cxj}, which can efficiently deal with massless propagator-type diagrams in $d=4-2\epsilon$ 
dimensional regularization~\cite{Bollini:1972ui, tHooft:1972tcz} 
up to four loops. 
In this way we can obtain fixed moments of the OMEs in Eq.~(\ref{eq:OMEs}). Finally, the calculations are done using a general covariant gauge. 
Although the operators considered are gauge invariant, the OMEs will in general depend on the gauge parameter. 
Using a general covariant gauge, the independence of the anomalous dimensions of the gauge parameter then provides a check on our calculations.

For the renormalization we use the \MSb-scheme \cite{tHooft:1973mfk, Bardeen:1978yd}. 
In this scheme, the evolution of the strong coupling is governed by
\begin{equation}
    \frac{da_s}{d\ln \mu^2} = \beta(a_s) =-a_s(\epsilon+\beta_0 a_s +\beta_1 a_s^2 +\beta_2 a_s^3 + \dots)
\end{equation}
with $\beta(a_s)$ the standard QCD beta-function and 
$\beta_0 = (11/3) \ca - (2/3) \nf$.
Writing out Eq.~(\ref{eq:renormPattern}), the operators are then renormalized as 
\begin{equation}
\label{eq:renorm}
    \mathcal{O}_{N+1} = Z_{\psi}\, 
    (Z_{N,N}\mathcal[{O}_{N+1}] + Z_{N,N-1}[\partial \mathcal{O}_{N}] + \dots  + Z_{N,0}[\partial^N \mathcal{O}_{1}])
    \, ,
\end{equation}
where the quark wave function renormalization factor $Z_{\psi}$ takes care of the self-energy corrections for the off-shell external quarks and we have abbreviated, cf.~Eq.~(\ref{eq:theDeltas}), 
\begin{equation}
\label{eq:ONdef}
    \Delta^{\mu_1}\dots \Delta^{\mu_N}\,
    \mathcal{O}^{NS}_{\mu_1\dots \mu_{N}} \,\equiv\, \mathcal{O}_{N}
    \, .
\end{equation}
In this notation, $\mathcal{O}_1$ simply represents the vector current
\begin{equation}
    \mathcal{O}_1 \,=\, 
    \Delta^{\mu}\, \overline{\psi}\lambda^{\alpha}\gamma_{\mu}\psi.
\end{equation}
Since this is a conserved quantity, its anomalous dimension vanishes and $Z_{0,0}=1$. 
For the full set of operators up to spin $(N+1)$, the renormalization takes the form of a matrix equation
\begin{eqnarray}
\label{eq:renormMatrix}
    \begin{pmatrix}
            \mathcal{O}_{N+1} \\ \partial \mathcal{O}_{N} \\ \vdots \\ 
            \partial^k \mathcal{O}_{N+1-k} \\ \vdots \\ \partial^N \mathcal{O}_{1}
    \end{pmatrix} 
    \,=\, Z_{\psi}  
    \begin{pmatrix}
            Z_{N,N} & Z_{N,N-1} & ... & Z_{N,N-k} & ... & Z_{N,0} \\
            0 & Z_{N-1,N-1} & ... & Z_{N-1,N-k} & ... & Z_{N-1,0} \\
            \vdots & \vdots & ... & \vdots & ... & \vdots \\
            0 &  0 & ... & Z_{N-k,N-k} & ... & Z_{N-k,0} \\
            \vdots & \vdots & ... &\vdots & ... & \vdots \\
            0 & 0 & ... & 0 & ... & 1
    \end{pmatrix} 
    \begin{pmatrix}
            [\mathcal{O}_{N+1}] \\ [\partial \mathcal{O}_{N}] \\ \vdots \\ 
            [\partial^k \mathcal{O}_{N+1-k}] 
            \\ \vdots \\ [\partial^N \mathcal{O}_{1}]
    \end{pmatrix}
    \, .
\end{eqnarray}
Note that the bottom-right $n \times n$ submatrix (with $n<N$) represents the full mixing matrix for the renormalization of the spin-$n$ operators. For example, the $2 \times 2$ submatrix of Eq.~(\ref{eq:renormMatrix})
\begin{equation}
    \begin{pmatrix}
            Z_{1,1} & Z_{1,0} \\
            0 & 1
    \end{pmatrix}
\end{equation}
is exactly the matrix appearing in the renormalization of the set of spin-2 operators $\{\mathcal{O}_2,\partial\mathcal{O}_1\}$.

The anomalous dimensions $\gamma_{N,k}^{\mathcal{D}}$ emerge from the $Z$-factors 
according to Eq.~(\ref{eq:gamZ}) and can be expanded in a power series in $a_s$
\begin{equation*}
    \gamma_{N,k}^{\mathcal{D}} \,=\, 
    a_s\gNknlo + a_s^2\gNknnlo + a_s^3\gNknnnlo + a_s^4\gNknnnnlo + a_s^5\gNknnnnnlo
    + \dots \, .
\end{equation*}
The explicit form of the $Z$-factors in a perturbative expansion in dimensional regularization in the \MSb-scheme can be obtained from Eq.~(\ref{eq:gamZ}) in terms of the anomalous dimensions and the coefficients of the QCD beta-function. 
For illustration we quote them up to order $a_s^3$, but the expansions can readily be generalized to higher orders. 
We present separately the diagonal factors $Z_{N,N}$, which 
are just those that renormalize the forward operators  
and the off-diagonal ones $Z_{N,k}$, where $k \neq N$ is understood.
\begin{eqnarray}
  \label{eq:3lZfactor}
{\lefteqn{
  Z_{N,N} \,=\, 
  1 + \frac{a_s}{\epsilon}\gamma_{N,N}^{(0)}
  + \frac{a_s^2}{2\epsilon}\Bigg\{\frac{1}{\epsilon}
  \Big(\gamma_{N,N}^{(0)}-\beta_0\Big)\gamma_{N,N}^{(0)}
  +\gamma_{N,N}^{(1)}\Bigg\}
  }}
\nonumber\\ &&
  + \frac{a_s^3}{6\epsilon}\Bigg\{\frac{1}{\epsilon^2}
  \Big(\gamma_{N,N}^{(0)}-\beta_0\Big) \Big(\gamma_{N,N}^{(0)}-2\beta_0\Big)\gamma_{N,N}^{(0)}
    +\frac{1}{\epsilon}\Big(3\gamma_{N,N}^{(0)}-2\beta_0\Big)\gamma_{N,N}^{(1)}
  -\frac{2}{\epsilon}\beta_1\gamma_{N,N}^{(0)}
  +2\gamma_{N,N}^{(2)} \Bigg\} 
  \, ,
  \qquad
%
\\[1ex]
  \label{eq:offd-3lZfactor}
{\lefteqn{
  Z_{N,k} \,=\, \frac{a_s}{\epsilon}\gNknlo
  + \frac{a_s^2}{2\epsilon}\Bigg\{\frac{1}{\epsilon}
  \Big(\gamma_{N,N}^{(0)}+\gamma_{k,k}^{(0)}-\beta_0\Big)\gNknlo
  +\frac{1}{\epsilon}\sum_{i=k+1}^{N-1}\gamma_{N,i}^{\mathcal{D}, (0)}\gamma_{i,k}^{\mathcal{D}, (0)}
  +\gNknnlo\Bigg\} 
}}
  \nonumber\\ &&
  +
  \frac{a_s^3}{6\epsilon}\Bigg\{
  \frac{2}{\epsilon^2}\beta_0^2\gNknlo
  -\frac{3}{\epsilon^2}\beta_0\Bigg(\gNknlo\Big(\gamma_{N,N}^{(0)}+\gamma_{k,k}^{(0)}\Big)
  +\sum_{i=k+1}^{N-1}\gamma_{N,i}^{\mathcal{D}, (0)}\gamma_{i,k}^{\mathcal{D}, (0)}\Bigg)
  \nonumber\\ &&
  \qquad
  +\frac{1}{\epsilon^2}\sum_{i=k}^{N}\sum_{j=1}^{i}
  \gamma_{N,i}^{\mathcal{D}, (0)}\gamma_{i,j}^{\mathcal{D}, (0)}\gamma_{j,k}^{\mathcal{D}, (0)}
  -\frac{2}{\epsilon}\beta_1\gNknlo
  -\frac{2}{\epsilon}\beta_0\gNknnlo
    \nonumber\\ &&
    \qquad
  +\frac{1}{\epsilon}\sum_{i=1}^{N}\Big(2\gamma_{N,i}^{\mathcal{D}, (0)}\gamma_{i,k}^{\mathcal{D}, (1)}+\gamma_{N,i}^{\mathcal{D}, (1)}\gamma_{i,k}^{\mathcal{D}, (0)}\Big)
  + 2\gNknnnlo \Bigg\}.
\end{eqnarray}
The mixing under renormalization is manifest in the appearance 
of sums over anomalous dimensions, starting at order $a_s^2$ 
in the expression for $Z_{N,k}$.

The computation of the bare OMEs for the operators $\mathcal{O}_{N+1}$ 
in Eq.~(\ref{eq:ONdef}) for fixed moments $N$ then allows to reconstruct 
the last column of the mixing matrix as we will describe now. 
At each order in perturbation theory one determines the quantity $\mathcal{B}(N+1)$ from the ($1/\epsilon)$-pole of the bare OME~\footnote{It is understood that coupling and gauge constants are already renormalized.} of the spin-($N+1$) operator. 
Using the $Z$-factors in Eqs.~(\ref{eq:3lZfactor}) and  (\ref{eq:offd-3lZfactor}) and the fact that the OMEs are renormalized according to Eq.~(\ref{eq:renorm}), the expression for $\mathcal{B}$ can be identified as the sum of the elements in the $N$-th row of the mixing matrix,
\begin{equation}
\mathcal{B}(N+1) \,=\, \sum_{j=0}^{N} \gamma_{N,j}^{\mathcal{D}}
    \, .
\end{equation}
The computation of the bare OMEs uses fixed moments and Eq.~(\ref{mainK0})
can be rewritten as a relation for the last column $\gamma_{N,0}^{\mathcal{D}}$,
\begin{equation}
    \gamma_{N,0}^{\mathcal{D}} \,=\, (-1)^N\Bigg\{\mathcal{B}(N+1)
    -\sum_{j=1}^{N-1}(-1)^j\binom{N}{j}\gamma_{j,0}^{\mathcal{D}}\Bigg\}
    \, ,
\end{equation}
which illustrates the bootstrap in $N$.
For a given moment $N$, the element $\gamma_{N,0}^{\mathcal{D}}$ of the last column is expressed in terms of the elements $\gamma_{j,0}^{\mathcal{D}}$ with $j<N$ and the sum of the elements in the $N$-th row of the mixing matrix.
Using in addition Eq.~(\ref{g0}) it is straightforward to rewrite this as a recursion in $\mathcal{B}$
\begin{equation}
\label{g0-from-B}
    \gamma_{N,0}^{\mathcal{D}} \,=\, \sum_{i=0}^{N}(-1)^i\binom{N}{i}\mathcal{B}(i+1)
    \,=\,
    {\mathcal{C}}\, \mathcal{B}(N+1)\, ,
\end{equation}
which is nothing but the conjugation operation, applied to $\mathcal{B}$.
The conjugation of Eq.~(\ref{g0-from-B}) leads then back to Eq.~(\ref{mainK0}) as the special case $k=0$ of Eq.~(\ref{mainGamma}) for constraints on the anomalous dimensions. One might expect similar relations to follow from Eq.~(\ref{mainGamma}) for different values of $k$. 
However, this not the case, since for $k \neq 0$ the corresponding expression does not involve the sum of the anomalous dimensions by itself, but instead a weighted version of this. For example, the $k=1$ relation involves
\begin{equation}
    \sum_{j=0}^{N}\, j\, \gamma_{N,j}^{\mathcal{D}}\, ,
\end{equation}
which cannot be directly related to Feynman diagram computations of fixed moments of the bare OMEs in Eq.~(\ref{eq:OMEs}).

\subsection{Calculating sums}
At the $l$-loop level, the forward anomalous dimensions in general consist of 
harmonic sums of maximum weight $w=2l-1$ and denominators in $N+\alpha$ (with $\alpha \in \mathbb{N}$) up to the same maximum power. More specifically, the maximum weight of a specific term in the anomalous dimension depends on both its color-structure and on the number of powers of $\nf$ multiplying it. Terms multiplied by the color factors $\ca=\nc$ and $\cf=(n_c^2-1)/(2\nc)$ can have weight up to $2l-1$, while each additional factor of $\nf$ will decrease this maximum weight by one, down to weight $l$ for the leading-$\nf$ terms. A similar reasoning holds for terms multiplied by values of the Riemann-zeta function $\zeta_n$ defined as
\begin{equation}
    \zeta_n = \sum_{i=1}^{\infty}i^{\, -n}.
\end{equation}
The maximum weight of a term containing $\zeta_n$ is in general $2l-1-n$, and $l-n$ for the leading-$\nf$ terms specifically.
Harmonic sums at argument $N$ are recursively defined by~\cite{Vermaseren:1998uu,Blumlein:1998if} 
\begin{eqnarray}
\label{eq:Hsum}
  S_{\pm m}(N) &=& \sum_{i=1}^{N}\; (\pm 1)^i \, i^{\, -m}
  \, , 
  \nonumber\\
  S_{\pm m_1^{},\,m_2^{},\,\ldots,\,m_d}(N) &=& \sum_{i=1}^{N}\:
  (\pm 1)^{i} \; i^{\, -m_1^{}}\; S_{m_2^{},\,\ldots,\,m_d}(i)
\, ,
\end{eqnarray}
where the weight $w$ is defined by the sum of the absolute
values of the indices~$m_d$.
Based on the  constraints in Eq.~(\ref{mainConj}) 
we expect a similar structure for the off-forward anomalous dimensions $\gamma_{N,k}^{\mathcal{D}, (l-1)}$ at $l$-loops. 

The constructive use of the conjugation in Eq.~(\ref{mainConj}) requires the evaluation of single and double sums over such structures. 
These sums are non-standard in the sense that they are outside the class of sums that can be solved by the algorithms encoded in the {\sc Summer} program~\cite{Vermaseren:1998uu} in {\sc Form}, which has been a standard in the calculation of the forward anomalous dimensions. 
However, the type of sums appearing can be dealt with using the principles of symbolic summation, 
see e.g. \cite{books/daglib/0068021,DBLP:series/tmsc/KauersP11} for extensive overviews. 
In particular, the {\sc Mathematica} package 
{\sc Sigma}~\cite{Schneider2004} is very helpful.
For this reason, we briefly review aspects of (creative) telescoping and the key features of {\sc Sigma}.

Suppose one wants to find a closed form for some summation
\begin{equation}
    \sum_{k=a}^N f(k)\, .
\end{equation}
Often it is possible to solve this problem using the telescoping algorithm. The task is then to find a function $g(N)$ such that the summand can be written as
\begin{equation}
    f(k) \,=\, \Delta g(k) \equiv g(k+1)-g(k)
    \, .
\end{equation}
Here $\Delta$ represents the finite-difference operator. 
Whenever this is possible, the summation problem can be solved as
\begin{equation}
    \sum_{k=a}^N f(k) \,=\, \sum_{k=a}^N g(k+1) - \sum_{k=a}^N g(k) 
    \,=\, g(N+1)-g(a)
    \, .
\end{equation}
A generalization of this algorithm to hypergeometric sums of the type
\begin{equation}
    \sum_{k=a}^b f(n,k) \, \equiv \, S(n)
\end{equation}
was constructed by Zeilberger and is called creative telescoping \cite{Zeilberger1991}. Here, the task is to find functions $c_0(n),\dots ,c_d(n)$ and $g(n,k)$ such that 
\begin{equation}
    g(n,k+1)-g(n,k) = c_0(n)f(n,k) + \dots  + c_d(n)f(n+d,k)
    \, .
\end{equation}
Summing both sides of the equation then gives, applying telescoping to the left-hand side
\begin{equation}
    g(n,b+1)-g(n,a) \,=\, 
    c_0(n) \sum_{k=a}^b f(n,k) + \dots  + c_d(n) \sum_{k=a}^b f(n+d,k)
    \, .
\end{equation}
Hence one gets an inhomogeneous recurrence for the original sum of the form
\begin{equation}
    q(n) \,=\, c_0(n) S(n) + \dots  + c_{d}(n) S(n+d)
    \, .
\end{equation}
The creative telescoping algorithm is the main method used in {\sc Sigma} to solve summation problems. 
After the recurrence is generated, {\sc Sigma} first solves this equation in terms of the solution of the homogeneous version and a particular solution. 
The final closed form expression for the initial sum is then given 
as a linear combination of the solutions of the recurrence that has the same initial values as the sum.

%
\section{Results up to five loops in the leading-$\nf$ limit}
\label{sec:Nres}
\setcounter{equation}{0}

Here we explore the large-$n_f$ limit of the mixing matrices 
up to five loops using the consistency relations for the anomalous dimension matrices discussed above along with explicit results 
for the forward anomalous dimensions~\cite{Gracey:1994nn,Davies:2016jie} 
and direct computations of the relevant OMEs in Eq.~(\ref{eq:OMEs}) up to four loops. Up to the three-loop level, the results for $\gamma_{N,k}^{\mathcal{G}}$
in the Gegenbauer basis are known~\cite{Braun:2017cih}. 
In the total derivative basis, only fixed low-$N$ results for
$\gamma_{N,k}^{\mathcal{D}}$ are available~\cite{Gracey:2009da,Kniehl:2020nhw}. 

\subsection{One-loop anomalous dimensions}
\label{1loopRES}
At the one-loop level, the mixing matrix $\hat{\gamma}_{N}^{\mathcal{G}}$ in the Gegenbauer basis is diagonal \cite{Efremov:1978rn, Makeenko:1980bh}, i.e.\footnote{Here and in the following, $\gamma_{N,k}$ is understood to mean $\gamma_{N,k} \text{\:with\:} k \neq N$.}
\begin{equation}
    \gamma_{N,k}^{\mathcal{G},{(0)}} \,=\, 0
    \, .
\end{equation}
This implies that the operators with spin $N=5$ just renormalize multiplicatively at the one-loop level with\footnote{To facilitate comparison between the two operator bases, we construct the mixing matrices in the Gegenbauer basis in upper triangular form, so that the bottom row corresponds to $N=1$. This differs from the conventions used in \cite{Braun:2017cih}. Additionally, an extra factor of $1/2$ appears as compared to \cite{Braun:2017cih} because of different conventions for the definition of the anomalous dimensions.}
\begin{equation}
    \hat{\gamma}^{\mathcal{G}, {(0)}}_{N=5} \,=\,  \cf \begin{pmatrix}
          \frac{91}{15} && 0 && 0 && 0 && 0 \\[6pt]
          0 && \frac{157}{30} && 0 && 0 && 0 \\[6pt]
          0 && 0 && \frac{25}{6} && 0 && 0 \\[6pt]
          0 && 0 && 0 && \frac{8}{3} && 0 \\[6pt]
          0 && 0 && 0 && 0 && 0 
    \end{pmatrix}
    \, .
\end{equation}
In contrast, in the total derivative basis the off-diagonal elements are non-zero already at this order. 
Application of the algorithm described in Sec.~\ref{sec:constraintsADM} leads to
\begin{equation}
\label{eq:1loopADM}
    \gamma_{N,k}^{\mathcal{D}, (0)} \,=\, 
    \cf\*\Big(\frac{2}{N+2}-\frac{2}{N-k}\Big)
    \, ,
\end{equation}
which is consistent with the result in the Gegenbauer basis according to Eq.~(\ref{eq:GtoD}). We also find agreement with the fixed moments presented in~\cite{Gracey:2009da}.
For spin-$5$ operators in the total derivative basis, the mixing matrix is then
\begin{equation}
   \hat{\gamma}^{\mathcal{D}, {(0)}}_{N=5} \,=\,  \cf \begin{pmatrix}
          \frac{91}{15} && -\frac{5}{3} && -\frac{2}{3} && -\frac{1}{3} && -\frac{1}{6} \\[6pt]
          0 && \frac{157}{30} && -\frac{8}{5} && -\frac{3}{5} && -\frac{4}{15} \\[6pt]
          0 && 0 && \frac{25}{6} && -\frac{3}{2} && -\frac{1}{2} \\[6pt]
          0 && 0 && 0 && \frac{8}{3} && -\frac{4}{3} \\[6pt]
          0 && 0 && 0 && 0 && 0 
    \end{pmatrix}
    \, .
\end{equation}

{
At this order, another check can be made, namely by comparing with the anomalous dimensions in the Geyer basis \cite{Blumlein:1999sc, Geyer:1982fk}. 
The off-diagonal piece of the mixing matrix at one-loop is~\cite{Blumlein:1999sc, Geyer:1982fk}\footnote{{We have an additional factor of -2 here compared to \cite{Blumlein:1999sc, Geyer:1982fk}, coming from the different conventions used in defining the anomalous dimensions.}}
\begin{equation}
    \gamma_{N,k}^{\mathcal{B}} = -2\cf\Bigg[\frac{1}{(N+1)(N+2)}+2\frac{k+1}{(N-k)(N+1)}\Bigg].
\end{equation}
Using Eq.~(\ref{eq:geyerRen}) for $k=N$ we find the following relation between the anomalous dimensions in the different operator bases
\begin{equation}
    \gamma_{N,N}^{(0)} + \sum_{j=0}^{N-1}\frac{1 \pm (-1)^j}{2}\gamma_{N,j}^{(0),\mathcal{B}} 
    = \pm (-1)^{N}\sum_{l=0}^{N}2^l(-1)^l\binom{N}{l}\gamma_{l,0}^{(0), \mathcal{D}}
    \, ,
\end{equation}
where the $'+'$-sign holds for even values of $N$ 
and the $'-'$-sign for odd $N$.
We have checked that our result Eq.~(\ref{eq:1loopADM}) obeys these relations.
For spin-$5$ operators in the Geyer basis, the one-loop mixing matrix is then
\begin{equation}
   \hat{\gamma}^{\mathcal{B}, {(0)}}_{N=5} \,=\,  \cf \begin{pmatrix}
          \frac{91}{15} && -\frac{49}{15} && -\frac{19}{15} && -\frac{3}{5} && -\frac{4}{15} \\[6pt]
          0 && \frac{157}{30} && -\frac{31}{10} && -\frac{11}{10} && -\frac{13}{30} \\[6pt]
          0 && 0 && \frac{25}{6} && -\frac{17}{6} && -\frac{5}{6} \\[6pt]
          0 && 0 && 0 && \frac{8}{3} && -\frac{7}{3} \\[6pt]
          0 && 0 && 0 && 0 && 0 
    \end{pmatrix}
    \, .
\end{equation}
}

\subsection{Two-loop anomalous dimensions}
Beyond the one-loop level, also the mixing matrix in the Gegenbauer basis gets non-zero off-diagonal contributions. These can be calculated in general as \cite{Mueller:1993hg, Braun:2017cih}
\begin{eqnarray}
  \label{eq:mueller}
  \hat{\gamma}^{\mathcal{G}}(a_s) \,=\,
  \textbf{G} \left\{ [\hat{\gamma}^{\mathcal{G}}(a_s),\hat{b}]\left(\frac12 \hat{\gamma}^{\mathcal{G}}(a_s)+\beta(a_s) \right)
    + [\hat{\gamma}^{\mathcal{G}}(a_s),\hat{w}(a_s)] \right\}
  \, ,
\end{eqnarray}
in terms of the matrix commutators denoted as $[\ast, \ast]$ and with
\begin{equation}
    \textbf{G}\{\hat{M}\}_{N,k} \,=\, -\frac{M_{N,k}}{a(N,k)}
    \, ,
\end{equation}
and
\begin{eqnarray}
    a(N,k) &=& (N-k)(N+k+3)\, ,\\
    \hat{b}_{N,k} &=& -2k\delta_{N,k}-2(2k+3)\vartheta_{N,k}
    \, .
\end{eqnarray}
The discrete step-function in the last term is defined as
\begin{equation}
\vartheta_{N,k} \,\equiv\,
    \begin{cases}
     1 \quad \text{if\:} N-k > 0 \text{\:and even} \\  0 \quad \text{else.}
    \end{cases}
\end{equation}
The $N-k$ even condition originates from the fact that, in the Gegenbauer basis, only CP-even operators are considered. The conformal anomaly, $\hat{w}(a_s)$, can be written as a power series in the strong coupling
\begin{equation}
    \hat{w}(a_s) \,=\, a_s\hat{w}^{(0)}+a_s^2\hat{w}^{(1)} + \dots \: .
\end{equation}
For the determination of the mixing matrix at order $a_s^{l}$, the conformal anomaly is only needed up to order $a_s^{l-1}$ \cite{Mueller:1991gd}.

At the two-loop level, the general relation Eq.~(\ref{eq:mueller}) leads to \cite{Braun:2017cih}
\begin{equation}
    \gamma_{N,k}^{\mathcal{G}, (1)} \,=\, \delta_{N,k}\gamma_{k,k}^{(1)} -\frac{\gamma_{N,N}^{(0)}-\gamma_{k,k}^{(0)}}{a(N,k)}\Bigg\{-2(2k+3)\Big(\beta_0+\frac{1}{2}\gamma_{k,k}^{(0)}\Big)\vartheta_{N,k}+w_{N,k}^{(0)}\Bigg\}
    \, .
\end{equation}
For the leading-$n_f$ contributions, only the term proportional to $\beta_0$ is relevant. The anomalous dimension can also be written in terms of harmonic sums and denominators, giving for the leading-$n_f$ piece
\begin{dmath}
\gamma_{N,k}^{\mathcal{G}, (1)} \,=\, \frac{8}{3}\frac{n_f C_F}{a(N,k)}\vartheta_{N,k}\Bigg\{-2\Big(S_{1}(N)-S_{1}(k)\Big)(2k+3)-(2k+3)\Big(\frac{1}{N+1}+\frac{1}{N+2}\Big)+4+\frac{1}{k+1}-\frac{1}{k+2} \Bigg\}
\end{dmath}.
%
For the spin-5 operators, the leading-$n_f$ piece of the mixing matrix is then
\begin{equation}
  \hat{\gamma}^{\mathcal{G}, {(1)}}_{N=5} \,=\,  -\nf \cf  \begin{pmatrix}
            \frac{7783}{1350} && 0 && \frac{133}{135} && 0 && \frac{13}{15} \\[6pt]
0 && \frac{13271}{2700} && 0 && \frac{11}{9} && 0 \\[6pt]
0 && 0 && \frac{415}{108} && 0 && \frac{5}{3} \\[6pt]
0 && 0 && 0 && \frac{64}{27} && 0 \\[6pt]
0 && 0 && 0 && 0 && 0 
    \end{pmatrix}
    \, .
\end{equation}
In the total derivative basis we find 
%
\begin{dmath}
\gamma_{N,k}^{\mathcal{D}, (1)} \,=\, \frac{4}{3}n_f C_F \Bigg\{\Big(S_{1}(N)-S_{1}(k)\Big)  \Big( \frac{1}{N+2} - \frac{1}{N-k}  \Big)
       + \frac{5}{3}\frac{1}{N-k} + \frac{2}{N+1} - \frac{11}{3}\frac{1}{N+2} + \frac{1}{(N+2)^2}\Bigg\}
\end{dmath}
and
\begin{equation}
    \hat{\gamma}^{\mathcal{D}, {(1)}}_{N=5} \,=\, \nf \cf \begin{pmatrix}
          -\frac{7783}{1350} && \frac{17}{10} && \frac{82}{135} && \frac{23}{90} && \frac{43}{540} \\[6pt]
          0 && -\frac{13271}{2700} && \frac{362}{225} && \frac{13}{25} && \frac{106}{675} \\[6pt]
          0 && 0 && -\frac{415}{108} && \frac{53}{36} && \frac{13}{36} \\[6pt]
          0 && 0 && 0 && -\frac{64}{27} && \frac{32}{27} \\[6pt]
          0 && 0 && 0 && 0 && 0 
    \end{pmatrix}
    \, .
\end{equation}
Using Eq.~(\ref{eq:GtoD}), we have checked that the expressions for $\gamma_{N,k}^{\mathcal{D}, (1)}$ and $\gamma_{N,k}^{\mathcal{G}, (1)}$ are consistent with one another. Furthermore, for the spin-2 and -3 operators, we find agreement with the results presented in \cite{Gracey:2009da}.

\subsection{Three-loop anomalous dimensions}
The Gegenbauer anomalous dimensions, using again Eq.~(\ref{eq:mueller}), are 
\begin{eqnarray}
\label{eq:3loopoff}
    \hat{\gamma}^{\mathcal{G}, (2)} &=& \textbf{G}\Bigg\{[\hat{\gamma}^{\mathcal{G}, (1)},\hat{b}]\Big(\frac{1}{2}\hat{\gamma}^{\mathcal{G}, (0)}+\beta_0\Big)+[\hat{\gamma}^{\mathcal{G}, (1)},\hat{w}^{(0)}]+[\hat{\gamma}^{\mathcal{G}, (0)},\hat{b}]\Big(\frac{1}{2}\hat{\gamma}^{\mathcal{G}, (1)}+\beta_1\Big) \nonumber \\ && +[\hat{\gamma}^{\mathcal{G}, (0)},\hat{w}^{(1)}]\Bigg\}
    \, .
\end{eqnarray}
An analytic expression for the two-loop conformal anomaly $\hat{w}^{(1)}$ 
depending on $N$ and $k$ in the space of Mellin moments 
is currently not available~\cite{Braun:2016qlg}.
Also, in the literature, e.g. \cite{Braun:2017cih}, only implicit relations of the form Eq.~(\ref{eq:3loopoff}) are given. The expansions in terms of harmonic sums here and at higher orders in the following sections are new.

Again, the leading-$n_f$ piece again just comes from the term proportional to $\beta_0$. In terms of harmonic sums we find
%
\begin{eqnarray}
\gamma^{\mathcal{G}, (2)}_{N,k} &=& \frac{32}{9}\frac{n_f^2 C_F}{a(N,k)}\vartheta_{N,k}\Bigg\{-\Big(S_{1}(N)-S_{1}(k)\Big)^2(2k+3)\nonumber \\&&+\Big(S_{1}(N)-S_{1}(k)\Big)(2k+3)\Big(\frac{5}{3}-\frac{1}{N+1}-\frac{1}{N+2}\Big)\nonumber \\&&+\Big(S_{1}(N)-S_{1}(k)\Big)\Big(4+\frac{1}{k+1}-\frac{1}{k+2}\Big)+(2k+3)\Big(-\frac{1}{6}\frac{1}{N+1}-\frac{1}{2}\frac{1}{(N+1)^2}\nonumber \\&&+\frac{11}{6}\frac{1}{N+2}-\frac{1}{2}\frac{1}{(N+2)^2}\Big)-\frac{10}{3}+\frac{2}{N+1}+\frac{1}{N+1}\frac{1}{k+1}-\frac{5}{6}\frac{1}{k+1}-\frac{1}{2}\frac{1}{(k+1)^2}\nonumber \\&&+\frac{2}{N+2}-\frac{1}{N+2}\frac{1}{k+2}+\frac{5}{6}\frac{1}{k+2}+\frac{1}{2}\frac{1}{(k+2)^2}\Bigg\}
\, ,
\end{eqnarray}
%
and explicitly for the spin-5 operators
\begin{equation}
   \hat{\gamma}^{\mathcal{G}, {(2)}}_{N=5} \,=\,  n_f^2 \cf \begin{pmatrix}
            -\frac{215621}{121500} && 0 && \frac{10339}{12150} && 0 && \frac{329}{1350} \\[6pt]
0 && -\frac{384277}{243000} && 0 && \frac{793}{810} && 0 \\[6pt]
0 && 0 && -\frac{2569}{1944} && 0 && \frac{59}{54} \\[6pt]
0 && 0 && 0 && -\frac{224}{243} && 0 \\[6pt]
0 && 0 && 0 && 0 && 0 
    \end{pmatrix}
    \, .
\end{equation}
In the total derivative basis our method gives
\begin{dmath}
\gamma^{\mathcal{D}, (2)}_{N,k} \,=\, \frac{4}{9} n_f^2 C_F \Bigg\{\Big(S_{1}(N)-S_{1}(k)\Big)^2  \Big(  \frac{1}{N+2} - \frac{1}{N-k} \Big)
       + 2\Big(S_{1}(N)-S_{1}(k)\Big)  \Big( \frac{5}{3}\frac{1}{N-k} + \frac{2}{N+1} - \frac{11}{3}\frac{1}{N+2} + \frac{1}{(N+2)^2}\Big)
       + \Big(S_{2}(N)-S_{2}(k)\Big)  \Big( \frac{1}{N+2} - \frac{1}{N-k} \Big)
       + \frac{2}{3}\frac{1}{N-k} - \frac{26}{3}\frac{1}{N+1} + \frac{4}{(N+1)^2} + \frac{8}{N+2} - \frac{22}{3}
         \frac{1}{(N+2)^2} + \frac{2}{(N+2)^3} \Bigg\}
    \, ,
\end{dmath}
and
\begin{equation}
    \hat{\gamma}^{\mathcal{D}, {(2)}}_{N=5} \,=\, n_f^2 \cf \begin{pmatrix}
          -\frac{215621}{121500}  && \frac{3131}{8100} && \frac{1312}{6075} && \frac{1181}{8100} && \frac{4841}{48600} \\[6pt]
          0 && -\frac{384277}{243000} && \frac{3947}{10125} && \frac{734}{3375} && \frac{4411}{30375} \\[6pt]
          0 && 0 && -\frac{2569}{1944} && \frac{259}{648} && \frac{151}{648} \\[6pt]
          0 && 0 && 0 && -\frac{224}{243} && \frac{112}{243} \\[6pt]
          0 && 0 && 0 && 0 && 0 
    \end{pmatrix}
    \, .
\end{equation}
Again the consistency relation between $\gamma^{\mathcal{D}, (2)}_{N,k}$ and $\gamma^{\mathcal{G}, (2)}_{N,k}$ in Eq.~(\ref{eq:GtoD}) checks out. 
Additionally, the result agrees with the analytical calculation of the $N=2$ matrix in \cite{Gracey:2009da} and a numerical calculation of the $N=3$ matrix in \cite{Kniehl:2020nhw}.

\subsection{Four-loop anomalous dimensions}
We start by presenting the results in the total derivative basis. There are two contributions to the anomalous dimensions, namely terms with and without a factor of $\zeta_3$.
Since $\zeta_3$ already has weight-3, the structure multiplying it will just be weight-1. Hence in complexity the contribution of such a term, using our algorithm, is equivalent to a one-loop calculation. We find
\begin{equation}
    \gamma^{\mathcal{D}, (3)}_{N,k}\biggr \vert_{\zeta_3} \,=\, \frac{32}{27} n_f^3 C_f \zeta_3\Big(\frac{1}{N+2}-\frac{1}{N-k}\Big)
    \, .
\end{equation}
and for the spin-5 operators
\begin{equation}
    \hat{\gamma}^{\mathcal{D}, {(3)}}_{N=5}\biggr \vert_{\zeta_3} \,=\, n_f^3 \cf \zeta_3 \begin{pmatrix}
         \frac{1456}{405}  && -\frac{80}{81} && -\frac{32}{81} && -\frac{16}{81} && -\frac{8}{81} \\[6pt]
          0 && \frac{1256}{405} && -\frac{128}{135} && -\frac{16}{45} && -\frac{64}{405} \\[6pt]
          0 && 0 && \frac{200}{81} && -\frac{8}{9} && -\frac{8}{27} \\[6pt]
          0 && 0 && 0 && \frac{128}{81} && -\frac{64}{81} \\[6pt]
          0 && 0 && 0 && 0 && 0 
    \end{pmatrix}
    \, .
\end{equation}
The expression for the $\zeta_3$-independent terms is
%
\begin{dmath}
\gamma^{\mathcal{D}, (3)}_{N,k} \,=\, \frac{8}{27}n_f^3 C_F \Bigg\{\frac{1}{3}\Big(S_{1}(N)-S_{1}(k)\Big)^3\Big(\frac{1}{N+2}-\frac{1}{N-k}\Big)+\Big(S_{1}(N)-S_{1}(k)\Big)^2\Big(\frac{5}{3}\frac{1}{N-k}+\frac{2}{N+1}-\frac{11}{3}\frac{1}{N+2}+\frac{1}{(N+2)^2}\Big)+\Big(S_{1}(N)-S_{1}(k)\Big)\Big(S_{2}(N)-S_{2}(k)\Big)\Big(\frac{1}{N+2}-\frac{1}{N-k}\Big)+2\Big(S_{1}(N)-S_{1}(k)\Big)\Big(\frac{1}{3}\frac{1}{N-k}-\frac{13}{3}\frac{1}{N+1}+\frac{2}{(N+1)^2}+\frac{4}{N+2}-\frac{11}{3}\frac{1}{(N+2)^2}+\frac{1}{(N+2)^3}\Big)+\Big(S_{2}(N)-S_{2}(k)\Big)\Big(\frac{5}{3}\frac{1}{N-k}+\frac{2}{N+1}-\frac{11}{3}\frac{1}{N+2}+\frac{1}{(N+2)^2}\Big)+\frac{2}{3}\Big(S_{3}(N)-S_{3}(k)\Big)\Big(\frac{1}{N+2}-\frac{1}{N-k}\Big)+\frac{2}{3}\frac{1}{N-k}+\frac{2}{N+1}-\frac{26}{3}\frac{1}{(N+1)^2}+\frac{4}{(N+1)^3}-\frac{8}{3}\frac{1}{N+2}+\frac{8}{(N+2)^2}-\frac{22}{3}\frac{1}{(N+2)^3}+\frac{2}{(N+2)^4} \Bigg\}
\, ,
\end{dmath}
leading to
\begin{equation}
    \hat{\gamma}^{\mathcal{D}, {(3)}}_{N=5} \,=\, n_f^3 \cf \begin{pmatrix}
          -\frac{10064827}{10935000}  && \frac{154397}{729000} && \frac{30097}{273375} && \frac{18049}{243000} && \frac{285967}{4374000} \\[6pt]
          0 && -\frac{17813699}{21870000} && \frac{191989}{911250} && \frac{32683}{303750} && \frac{237557}{2733750} \\[6pt]
          0 && 0 && -\frac{23587}{34992} && \frac{2401}{11664} && \frac{1477}{11664} \\[6pt]
          0 && 0 && 0 && -\frac{1024}{2187} && \frac{512}{2187} \\[6pt]
          0 && 0 && 0 && 0 && 0 
    \end{pmatrix}
    \, .
\end{equation}
Next we turn to the results in the Gegenbauer basis. These cannot be found in the literature, but can be calculated in the same way as the lower-order results. Since the one-loop beta-function and leading-$n_f$ three-loop forward anomalous dimensions do not depend on $\zeta_3$, there are no $\zeta_3$-terms in the off-diagonal part of the Gegenbauer mixing matrix
\begin{equation}
    \gamma_{N,k}^{\mathcal{G}, (3)}\biggr \vert_{\zeta_3} \,=\, 0
    \, ,
\end{equation}
which again parallels the one-loop case. Hence for $N=5$ we just have
\begin{equation}
\hat{\gamma}^{\mathcal{G}, {(3)}}_{N=5} \biggr \vert_{\zeta_3} \,=\, n_f^3 \cf \zeta_3 \begin{pmatrix}
      \frac{1456}{405}  && 0 && 0 && 0 && 0 \\[6pt]
          0 && \frac{1256}{405} && 0 && 0 && 0 \\[6pt]
          0 && 0 && \frac{200}{81} && 0 && 0 \\[6pt]
          0 && 0 && 0 && \frac{128}{81} && 0 \\[6pt]
          0 && 0 && 0 && 0 && 0 
\end{pmatrix}
    \, .
\end{equation}
For the $\zeta_3$-independent terms we find
%
\begin{dmath}
\gamma^{\mathcal{G}, (3)}_{N,k} \,=\, \frac{64}{27}\frac{n_f^3 C_F}{a(N,k)}\vartheta_{N,k}\Bigg\{-\frac{2}{3}\Big(S_{1}(N)-S_{1}(k)\Big)^3(2k+3)+\Big(S_{1}(N)-S_{1}(k)\Big)^2(2k+3)\Big(\frac{5}{3}-\frac{1}{N+1}-\frac{1}{N+2}\Big)+\Big(S_{1}(N)-S_{1}(k)\Big)^2\Big(4+\frac{1}{k+1}-\frac{1}{k+2}\Big)+\Big(S_{1}(N)-S_{1}(k)\Big)(2k+3)\Big(\frac{1}{3}-\frac{1}{3}\frac{1}{N+1}-\frac{1}{(N+1)^2}+\frac{11}{3}\frac{1}{N+2}-\frac{1}{(N+2)^2}\Big)+\Big(S_{1}(N)-S_{1}(k)\Big)\Big(-\frac{20}{3}+\frac{4}{N+1}+\frac{2}{N+1}\frac{1}{k+1}-\frac{5}{3}\frac{1}{k+1}-\frac{1}{(k+1)^2}+\frac{4}{N+2}-\frac{2}{N+2}\frac{1}{k+2}+\frac{5}{3}\frac{1}{k+2}+\frac{1}{(k+2)^2}\Big)+(2k+3)\Big(\frac{7}{3}\frac{1}{N+1}-\frac{1}{6}\frac{1}{(N+1)^2}-\frac{1}{2}\frac{1}{(N+1)^3}-\frac{2}{N+2}+\frac{11}{6}\frac{1}{(N+2)^2}-\frac{1}{2}\frac{1}{(N+2)^3}-\frac{1}{3}\Big(S_{3}(N)-S_{3}(k)\Big)\Big)-\frac{2}{3}+\frac{2}{3}\frac{1}{N+1}+\frac{2}{(N+1)^2}+\frac{1}{(N+1)^2}\frac{1}{k+1}-\frac{5}{3}\frac{1}{N+1}\frac{1}{k+1}-\frac{1}{N+1}\frac{1}{(k+1)^2}-\frac{8}{3}\frac{1}{k+1}+\frac{5}{6}\frac{1}{(k+1)^2}+\frac{1}{2}\frac{1}{(k+1)^3}-\frac{22}{3}\frac{1}{N+2}+\frac{2}{(N+2)^2}-\frac{1}{(N+2)^2}\frac{1}{k+2}+\frac{5}{3}\frac{1}{N+2}\frac{1}{k+2}+\frac{1}{N+2}\frac{1}{(k+2)^2}+\frac{5}{3}\frac{1}{k+2}-\frac{5}{6}\frac{1}{(k+2)^2}-\frac{1}{2}\frac{1}{(k+2)^3}\Bigg\}
\, ,
\end{dmath}
and
\begin{equation}
 \hat{\gamma}^{\mathcal{G}, {(3)}}_{N=5} = n_f^3 \cf \begin{pmatrix}
            -\frac{10064827}{10935000} && 0 && \frac{394793}{1093500} && 0 && \frac{65923}{121500} \\[6pt]
0 && -\frac{17813699}{21870000} && 0 && \frac{36491}{72900} && 0 \\[6pt]
0 && 0 && -\frac{23587}{34992} && 0 && \frac{797}{972} \\[6pt]
0 && 0 && 0 && -\frac{1024}{2187} && 0 \\[6pt]
0 && 0 && 0 && 0 && 0
    \end{pmatrix}
    \, .
\end{equation}
Again the consistency of the results for $\gamma^{\mathcal{G}, (3)}_{N,k}$ and $\gamma^{\mathcal{D}, (3)}_{N,k}$ 
based on Eq.~(\ref{eq:GtoD}) was explicitly checked.

\subsection{Discussion of the results in the total derivative basis}
From the structure of the anomalous dimensions $\gamma^{\mathcal{D}}_{N,k}$ presented here, we can make general predictions for the $\zeta$-independent terms of $\gamma^{\mathcal{D},(l)}_{N,k}$.
\begin{enumerate}
    \item The prefactor of $\frac{1}{(N+2)^{l}}$ can be written as
    \begin{equation}
       2 \cf (-\beta_0)^{l-1}\, , 
    \end{equation}
    where we take 
    \begin{equation}
        -\beta_0 \,=\, \frac{2}{3} n_f\, .
    \end{equation}
    \item The prefactor of the difference $S_{l-1}(N)-S_{l-1}(k)$ can be written as
    \begin{equation}
        \frac{(-\beta_0)^{l-1}}{l-1}\gamma_{N,k}^{\mathcal{D}, (0)} \: \: (l>1)\, .
    \end{equation}
    \item The prefactor of powers of $[S_1(N)-S_1(k)]^{l-1}$ can be written as
    \begin{equation}
        \frac{(-\beta_0)^{l-1}}{(l-1)!}\gamma_{N,k}^{\mathcal{D}, (0)} \: \: (l>1)\, .
    \end{equation}
    \item The term linear in the difference $S_1(N)-S_1(k)$ is just the term 
    which does not depend on harmonic sums at order $l-1$ multiplied by the factor $-\beta_0[S_1(N)-S_1(k)]$. 
    In general, we can write the prefactor of the $[S_1(N)-S_1(k)]^{\alpha}$ term as
    \begin{equation}
        \frac{-\beta_0}{\alpha}\gamma_{N,k}^{\mathcal{D}, (l-1)}\biggr \vert_{[S_1(N)-S_1(k)]^{\alpha-1}} \: \: (\alpha>0)\, .
    \end{equation}
    This also holds for products of such differences, e.g. the $[S_2(N)-S_2(k)][S_1(N)-S_1(k)]$ at the four-loop level corresponds to $-\beta_0[S_2(N)-S_2(k)]$ coming from the three-loop level. At this point, there is not enough information to determine the pattern for $[S_{\alpha}(N)-S_{\alpha}(k)]$ for $\alpha \neq 1$ and $\alpha \neq l-1$. 
    \item We can also say something about powers of denominators. If at the $l$-loop level we have a structure
    \begin{equation}
        \frac{1}{(N+\alpha)^{\delta}}
    \end{equation}
    then the $l+1$ expression will have the term
    \begin{equation}
        \frac{-\beta_0}{(N+\alpha)^{\delta+1}}\, ,
    \end{equation}
    where $\alpha=1,2$ and $\delta \neq l$.
\end{enumerate}
The above considerations imply that the $l$-loop expression for the leading-$n_f$ off-diagonal piece of the mixing matrix only has a small number unknowns, namely the coefficients of the weight $w=1$ terms and $[S_{\alpha}(N)-S_{\alpha}(k)]$. 
Also at higher loops we could expect terms of the form $[S_{\alpha}(N)-S_{\alpha}(k)]^{\beta}$ with $\alpha, \beta > 1$. 
The other higher-weight terms can be reconstructed from the lower-loop expressions. 
Hence, by considering just a few low $(N,k)$-pairs, it should be possible to use the conjugation relation, Eq.~(\ref{mainConj}), to completely fix the off-diagonal part of the mixing matrix. When combined with the all-order expression for the leading-$\nf$ terms of the forward anomalous dimensions, which was calculated in~\cite{Gracey:1994nn} using the critical exponents of the Wilson operators, this implies that the $\zeta$-independent terms of the leading-$\nf$ piece of the full mixing matrix can in principle be reconstructed to all orders. 

\subsubsection{Five-loop anomalous dimensions in the total derivative basis}
As an application of the discussion presented in the previous section, we determine the part of the leading-$\nf$ terms of the five-loop mixing matrix, which is independent of Riemann zeta-values.
For this, the five-loop expression for $\gamma_{N,N}$ presented in~\cite{Gracey:1994nn} is also used. In total there are six terms with a priori unknown numerical prefactors, namely
\begin{align}
    \Bigg\{&\frac{1}{N+2}, 
    \frac{1}{N+1}, 
    \frac{1}{N-k}, \nonumber\\& 
    [S_2(N)-S_2(k)]\Big(\frac{1}{3}\frac{1}{N-k}-\frac{13}{3}\frac{1}{N+1}+\frac{2}{(N+1)^2}+\frac{4}{N+2}-\frac{11}{3}\frac{1}{(N+2)^2}+\frac{1}{(N+2)^3}\Big), \nonumber\\& 
    [S_3(N)-S_3(k)]\Big(\frac{5}{3}\frac{1}{N-k}+\frac{2}{N+1}-\frac{11}{3}\frac{1}{N+2}+\frac{1}{(N+2)^2}\Big), \nonumber\\&
    [S_2(N)-S_2(k)]^2\Big(\frac{1}{N+2}-\frac{1}{N-k}\Big)\Bigg\}
    \, .
\end{align}
These can be fixed by using Eq.~(\ref{mainConj}) for $(N,k)$-pairs up to $N=4$, and we quickly find
%
\begin{eqnarray}
\gamma^{\mathcal{D}, (4)}_{N,k} &=& \frac{16}{81}n_f^4 C_F\Bigg\{\frac{1}{12}\Big(S_{1}(N)-S_{1}(k)\Big)^4\Big(\frac{1}{N+2}-\frac{1}{N-k}\Big)\nonumber\\&&+\frac{1}{3}\Big(S_{1}(N)-S_{1}(k)\Big)^3\Big(\frac{5}{3}\frac{1}{N-k}+\frac{2}{N+1}-\frac{11}{3}\frac{1}{N+2}+\frac{1}{(N+2)^2}\Big)\nonumber\\&&+\frac{1}{2}\Big(S_{1}(N)-S_{1}(k)\Big)^2\Big(S_{2}(N)-S_{2}(k)\Big)\Big(\frac{1}{N+2}-\frac{1}{N-k}\Big)\nonumber\\&&+\Big(S_{1}(N)-S_{1}(k)\Big)^2\Big(\frac{1}{3}\frac{1}{N-k}-\frac{13}{3}\frac{1}{N+1}+\frac{2}{(N+1)^2}+\frac{4}{N+2}\nonumber\\&&-\frac{11}{3}\frac{1}{(N+2)^2}+\frac{1}{(N+2)^3}\Big)\nonumber\\&&+\Big(S_{1}(N)-S_{1}(k)\Big)\Big(S_{2}(N)-S_{2}(k)\Big)\Big(\frac{5}{3}\frac{1}{N-k}+\frac{2}{N+1}-\frac{11}{3}\frac{1}{N+2}+\frac{1}{(N+2)^2}\Big)\nonumber\\&&+\frac{2}{3}\Big(S_{1}(N)-S_{1}(k)\Big)\Big(S_{3}(N)-S_{3}(k)\Big)\Big(\frac{1}{N+2}-\frac{1}{N-k}\Big)\nonumber\\&&+\Big(S_{1}(N)-S_{1}(k)\Big)\Big(\frac{2}{3}\frac{1}{N-k}+\frac{2}{N+1}-\frac{26}{3}\frac{1}{(N+1)^2}+\frac{4}{(N+1)^3}-\frac{8}{3}\frac{1}{N+2}\nonumber\\&&+\frac{8}{(N+2)^2}-\frac{22}{3}\frac{1}{(N+2)^3}+\frac{2}{(N+2)^4}\Big)+\frac{1}{4}\Big(S_{2}(N)-S_{2}(k)\Big)^2\Big(\frac{1}{N+2}-\frac{1}{N-k}\Big)\nonumber\\&&+\Big(S_{2}(N)-S_{2}(k)\Big)\Big(\frac{1}{3}\frac{1}{N-k}-\frac{13}{3}\frac{1}{N+1}+\frac{2}{(N+1)^2}+\frac{4}{N+2}\nonumber\\&&-\frac{11}{3}\frac{1}{(N+2)^2}+\frac{1}{(N+2)^3}\Big)+\frac{2}{3}\Big(S_{3}(N)-S_{3}(k)\Big)\Big(\frac{5}{3}\frac{1}{N-k}+\frac{2}{N+1}\nonumber\\&&-\frac{11}{3}\frac{1}{N+2}+\frac{1}{(N+2)^2}\Big)+\frac{1}{2}\Big(S_{4}(N)-S_{4}(k)\Big)\Big(\frac{1}{N+2}-\frac{1}{N-k}\Big)+\frac{2}{3}\frac{1}{N-k}\nonumber\\&&-\frac{2}{3}\frac{1}{N+1}+\frac{2}{(N+1)^2}-\frac{26}{3}\frac{1}{(N+1)^3}+\frac{4}{(N+1)^4}-\frac{8}{3}\frac{1}{(N+2)^2}+\frac{8}{(N+2)^3}\nonumber\\&&-\frac{22}{3}\frac{1}{(N+2)^4}+\frac{2}{(N+2)^5} \Bigg\}
\, .
\end{eqnarray}
%
 We have checked that the resulting expression obeys Eq.~(\ref{mainConj}) also for high moments. This then suggests the following pattern for the terms of the form 
$[S_{\alpha}(N)-S_{\alpha}(k)]$ at $l$ loops:
\begin{equation}
    (-\beta_0)\Big(\frac{2}{3}\Big)^{l+\alpha-3} \Tilde{\gamma}_{N,k}^{(l-1)} \biggr \vert_{[S_{\alpha-1}(N)-S_{\alpha-1}(k)]} \qquad  (\alpha \neq 1,l-1)
\end{equation}
where $\Tilde{\gamma}$ denotes that we only use the denominator-structure itself, i.e. we do not include the overall numerical factor of the lower-loop expression. The five-loop $N=5$ mixing matrix is
\begin{equation}
    \hat{\gamma}^{\mathcal{D}, {(4)}}_{N=5} = n_f^4 \cf \begin{pmatrix}
          -\frac{562208549}{984150000}  && \frac{8899139}{65610000} && \frac{1705289}{24603750} && \frac{2955389}{65610000} && \frac{15359729}{393660000} \\[6pt]
          0 && -\frac{990930013}{1968300000} && \frac{11153243}{82012500} && \frac{1831571}{27337500} && \frac{12500059}{246037500} \\[6pt]
          0 && 0 && -\frac{259993}{629856} && \frac{27955}{209952} && \frac{15439}{209952} \\[6pt]
          0 && 0 && 0 && -\frac{5504}{19683} && \frac{2752}{19683} \\[6pt]
          0 && 0 && 0 && 0 && 0 
    \end{pmatrix}
    \, .
\end{equation}

%
\section{Going beyond the leading-$\nf$-limit}
\label{sec:beyondNF}
\setcounter{equation}{0}
%
So far, the focus has been to apply our method to determine the mixing matrix in the limit of large $\nf$. However, the algorithm can also be applied beyond this limit, and we will present two possible threads of this. In the first the two-loop mixing matrix is determined in the planar limit, i.e. large $\nc$. 
In the second we present some low-$N$ matrices in full QCD.

\subsection{Second order mixing matrix in the leading color limit}
The application of our method to the determination of the leading-$n_f$ part of the mixing matrix works particularly well because of the simple structure of these terms. This simplicity manifests itself in three ways: (1) only harmonic sums with positive indices appear, 
(2) the number of possible terms is restricted because of the low maximum weight and, 
(3) increasing the order in $a_s$ by one corresponds to increasing the maximum weight of the possible structures by one. 
Moving away from the leading-$n_f$ limit and towards the leading color limit, point (1) still remains valid. 
Hence, while the sums that need to be evaluated according to our algorithm become more numerous, they are of the same type as those in the leading-$n_f$ limit, and the method goes through in exactly the same way. 

As an application of this, we present the off-diagonal piece of the second order mixing matrix in the planar limit. 
The two-loop anomalous dimensions can be written as
\begin{equation}
    \gamma_{N,k}^{\mathcal{D}, (1)} \,=\, \gamma_{N,k}^{\mathcal{D}, (1)}\biggr\vert_{n_f} +  \gamma_{N,k}^{\mathcal{D}, (1)}\biggr\vert_{\rm lc} + \gamma_{N,k}^{\mathcal{D}, (1)}\biggr\vert_{\rm nlc}
    \, .
\end{equation}
The first term just represents the leading-$\nf$ piece. The leading color term, $\gamma_{N,k}^{\mathcal{D}, (1)}\bigr\vert_{\rm lc}$, is proportional to $\cf \nc/2$ and the next-to-leading color one, $\gamma_{N,k}^{\mathcal{D}, (1)}\bigr\vert_{\rm nlc}$, to $\cf/(2\nc)$.

Using our method we then find
\begin{eqnarray}
    \gamma_{N,k}^{\mathcal{D}, (1)}\biggr\vert_{\rm lc} &=& 4\frac{\cf \nc}{2}\Bigg\{\frac{\Big(S_{1}(N)-S_{1}(k)\Big)^2}{N-k}\nonumber\\&&+\Big(S_{1}(N)-S_{1}(k)\Big)\Big(S_{1}(N-k)-S_{1}(k)\Big)\Big(\frac{1}{N+2} -\frac{2}{N-k}\Big)\nonumber\\&&+\frac{1}{2}\Big(S_{1}(N)-S_{1}(k)\Big)\Big(\frac{13}{3}\frac{1}{N-k}+\frac{4}{(N-k)^2}-\frac{2}{N+1}\frac{1}{k+1}  +\frac{2}{k+1}\frac{1}{N-k}\nonumber\\&&-\frac{13}{3}\frac{1}{N+2}-\frac{2}{(N+2)^2}\Big)+\Big(S_{1}(N-k)-S_{1}(k)\Big)\Big(\frac{1}{N+1}  +\frac{1}{N+1}\frac{1}{k+1}\nonumber\\&&-\frac{1}{k+1}\frac{1}{N+2}-\frac{1}{N+2}+\frac{1}{(N+2)^2}\Big) -\Big(S_{2}(N)-S_{2}(k)\Big)\Big(\frac{1}{N+2}-\frac{1}{N-k}\Big)\nonumber\\&&-2S_{2}(k)\Big(\frac{1}{N+2}-\frac{1}{N-k}\Big) -\frac{67}{9}\frac{1}{N-k}  -\frac{53}{6}\frac{1}{N+1}-\frac{1}{(N+1)^2}\nonumber\\&&-\frac{1}{(N+1)^2}\frac{1}{k+1}-\frac{1}{2}\frac{1}{N+1}\frac{1}{k+1}  +\frac{1}{2}\frac{1}{k+1}\frac{1}{N+2}  +\frac{293}{18}\frac{1}{N+2}-\frac{5}{3}\frac{1}{(N+2)^2}\nonumber\\&&-\frac{1}{(N+2)^3}-\frac{1}{(N+2)^2}S_{1}(k) \Bigg\}
    \, .
\end{eqnarray}
For illustration, we quote the corresponding mixing matrix for spin-5 operators
\begin{equation}
    \hat{\gamma}^{\mathcal{D}, {(1)}}_{N=5}\biggr\vert_{\rm lc} = \frac{\cf \nc}{2} \begin{pmatrix}
         \frac{654173}{13500}  && - \frac{2551}{216} && - \frac{7999}{1800} && - \frac{1439}{600} && - \frac{511}{300} \\[6pt]
          0 && \frac{255313}{6000} && - \frac{12881}{1125} && - \frac{39133}{9000} && - \frac{2936}{1125} \\[6pt]
          0 && 0 && \frac{15085}{432} && - \frac{1615}{144} && - \frac{673}{144} \\[6pt]
          0 && 0 && 0 && \frac{640}{27} && - \frac{320}{27} \\[6pt]
          0 && 0 && 0 && 0 && 0 
    \end{pmatrix}.
\end{equation}
The algorithm should also be applicable to the subleading color part $\gamma_{N,k}^{\mathcal{D}, (1)}\bigr\vert_{\rm nlc}$. 
However, because of the non-planar Feynman diagrams contributing,  harmonic sums with negative indices appear in the expression for the anomalous dimension. This adds an additional layer of complexity to the problem and we leave this part to future studies.

\subsection{$N=5$ matrix in full QCD}
Here we present the mixing matrix for spin-5 operators
\begin{equation}
    \begin{pmatrix}
            \gamma_{4,4} & \gamma_{4,3} & \gamma_{4,2} & \gamma_{4,1} & \gamma_{4,0} \\
            0 & \gamma_{3,3} & \gamma_{3,2} & \gamma_{3,1} & \gamma_{3,0} \\
            0 & 0 & \gamma_{2,2} & \gamma_{2,1} & \gamma_{2,0} \\
            0 & 0 & 0 & \gamma_{1,1} & \gamma_{1,0} \\
            0 & 0 & 0 & 0 & 0
    \end{pmatrix}
    \, ,
\end{equation}
in full QCD\footnote{At the three-loop level and beyond, we omit contributions of different non-singlet flavor structures proportional to $d^{abc}d_{abc}/n_c$; see e.g. \cite{Moch:2004pa} for details.}, i.e. $\cf = 4/3$ and $\ca = 3$. 
The corresponding expressions for a general gauge group $SU(\nc)$ can be found in Appendix~\ref{sec:appA}. 
Before presenting the explicit results, the relevant relations used in their derivation are reviewed.

\subsubsection{Relevant relations}
As explained in Sec.~\ref{sec:constraintsADM}, the next-to-diagonal elements can be calculated directly from the forward anomalous dimensions using Eq.~(\ref{NTD}).
This already fixes $\gamma_{1,0}$, $\gamma_{2,1}$, $\gamma_{3,2}$ and $\gamma_{4,3}$. For the last column, the relation between the Gegenbauer basis and the total derivative one, Eq.~(\ref{eq:GtoD}), can be used to determine $\gamma_{2,0}$, $\gamma_{3,0}$ and $\gamma_{4,0}$. 
Explicitly
\begin{align}
    \gamma_{2,0}^{\mathcal{D}} &= \frac{1}{30}\Big(\gamma_{2,0}^{\mathcal{G}}+6\gamma_{2,2}+30\gamma_{1,0}^{\mathcal{D}}\Big) \nonumber\\
    \gamma_{3,0}^{\mathcal{D}} &= \frac{1}{140}\Big(-3\gamma_{3,1}^{\mathcal{G}}-10\gamma_{3,3}-90\gamma_{1,0}^{\mathcal{D}}+210\gamma_{2,0}^{\mathcal{D}}\Big) \nonumber\\
    \gamma_{4,0}^{\mathcal{D}} &= \frac{1}{630}\Big(\gamma_{4,0}^{\mathcal{G}}+6\gamma_{4,2}^{\mathcal{G}}+15\gamma_{4,4}+210\gamma_{1,0}^{\mathcal{D}}-840\gamma_{2,0}^{\mathcal{D}}+1260\gamma_{3,0}^{\mathcal{D}}\Big).
\end{align}
Finally, we can use Eq.~(\ref{mainK0}) to express $\gamma_{3,1}^{\mathcal{D}}$ as
\begin{equation}
    \gamma_{3,1}^{\mathcal{D}} = 3\gamma_{2,0}^{\mathcal{D}}-3\gamma_{1,0}^{\mathcal{D}}-2\gamma_{3,0}^{\mathcal{D}}-\gamma_{3,2}^{\mathcal{D}}-\gamma_{3,3}.
\end{equation}
At $N=5$, there is not enough information to completely fix all the anomalous dimensions. Nevertheless, we can again use Eq.~(\ref{mainK0}) to derive
\begin{equation}
    \gamma_{4,1}^{\mathcal{D}}+\gamma_{4,2}^{\mathcal{D}} = -4\gamma_{1,0}^{\mathcal{D}}+6\gamma_{2,0}^{\mathcal{D}}-4\gamma_{3,0}^{\mathcal{D}}-\gamma_{4,3}^{\mathcal{D}}-\gamma_{4,4}
    \, .
\end{equation}
These relations can be used to determine the complete $N=2$ matrix up to the five-loop level. To the same order we can also calculate $\gamma_{2,1}$. Finally, with the three-loop information in the Gegenbauer basis from \cite{Braun:2017cih}, the full $N=4$ matrix can be calculated up to third order. For the fixed-$N$ forward anomalous dimensions we use the four-loop expressions from~\cite{Velizhanin:2011es,Velizhanin:2014fua,Ruijl:2016pkm,Moch:2017uml} 
and the five-loops ones from~\cite{Herzog:2018kwj}.
At the three-loop level, only the leading-$\nf$ term of $\gamma_{4,2}^{\mathcal{D}, (2)}$ is available. The remaining terms will be collected into $\Tilde{\gamma}_{4,2}^{\mathcal{D}, (2)}$.

\subsubsection{Results}
\label{sec:ResultsBeyondNf}
\begin{eqnarray}
    \gamma_{4,4} &=& 8.08889 \: a_s+(98.1994 - 7.68691 \: n_f) \: a_s^2\nonumber \\&&+(1720.94 - 279.180 \: n_f - 2.36621 \: n_f^2)\: a_s^3 \nonumber \\&& +(31273.3 - 7626.41 \: n_f + 277.106 \: n_f^2 + 4.53473 \: n_f^3)\: a_s^4+O(a_s^5) \\
    \gamma_{4,3}^{\mathcal{D}} &=& -2.22222\: a_s+(-23.8255 + 2.26667 \: n_f)\: a_s^2 \nonumber \\&& +(-410.760 + 68.9025 \: n_f + 0.515391 \: n_f^2)\: a_s^3 \nonumber \\&& +(-6915.76 + 1844.48 \: n_f - 76.8389 \: n_f^2 - 1.30057 \: n_f^3)\: a_s^4+O(a_s^5) \\
    \gamma_{4,2}^{\mathcal{D}} &=& -0.888889\: a_s+\Bigg(-8.88778+\gamma_{4,2}^{\mathcal{D}, (1)}\biggr\vert_{\rm nlc}+0.809877 \: n_f\Bigg)\: a_s^2 \nonumber \\&& +\Bigg(\Tilde{\gamma}_{4,2}^{\mathcal{D}, (2)} + 0.287956 \: n_f^2\Bigg)\: a_s^3+O(a_s^4) \\
    \gamma_{4,1}^{\mathcal{D}} &=& -0.444444\: a_s + \Bigg(-4.46178 - \gamma_{4,2}^{\mathcal{D}, (1)}\biggr\vert_{\rm nlc} +0.340741 \: n_f \Bigg)\: a_s^2 \nonumber \\&& +\Bigg(-234.566 + 40.5252 \: n_f + 0.194403 \: n_f^2 - \Tilde{\gamma}_{4,2}^{\mathcal{D}, (2)}\Bigg)\: a_s^3+O(a_s^4) \\
    \gamma_{4,0}^{\mathcal{D}} &=& -0.222222\: a_s + (-3.41128 + 0.106173 \: n_f)\: a_s^2 \nonumber \\&& + (-61.8523 + 7.90849 \: n_f + 0.132812 \: n_f^2)\: a_s^3+O(a_s^4) \\
    \gamma_{3,3} &=& 6.97778\: a_s + (86.2867 - 6.55358 \: n_f)\: a_s^2 \nonumber \\&& + (1515.56 - 244.729 \: n_f - 2.10852 \: n_f^2)\: a_s^3 \nonumber \\&& + (27815.4 - 6704.17 \: n_f + 238.687 \: n_f^2 + 3.88444 \: n_f^3)\: a_s^4 + O(a_s^5) \\
    \gamma_{3,2}^{\mathcal{D}} &=& -2.13333\: a_s + (-23.1028 + 2.14519 \: n_f)\: a_s^2 \nonumber \\&& + (-405.973 + 67.6368 \: n_f + 0.51977 \: n_f^2)\: a_s^3 \nonumber \\&& +(-7071.4 + 1807.36 \: n_f - 74.6209 \: n_f^2 - 1.23872 \: n_f^3)\: a_s^4 + O(a_s^5) \\
    \gamma_{3,1}^{\mathcal{D}} &=& -0.8\: a_s + (-8.66775 + 0.693333 \: n_f)\: a_s^2 \nonumber \\&&+(-142.057 + 25.4966 \: n_f + 0.289975 \: n_f^2)\: a_s^3  +O(a_s^4) \\
    \gamma_{3,0}^{\mathcal{D}} &=& -0.355556\: a_s + (-5.27039 + 0.209383 \: n_f)\: a_s^2 \nonumber \\&& + ( -95.1612 + 12.9537 \: n_f + 0.193624 \: n_f^2)\: a_s^3  + O(a_s^4) \\
    \gamma_{2,2} &=& 5.55556\: a_s + (70.8848 - 5.12346 \: n_f)\: a_s^2 \nonumber \\&& + (1244.91 - 199.637 \: n_f - 1.762 \: n_f^2)\: a_s^3 \nonumber \\&& + (23101.2 - 5499.27 \: n_f + 188.94 \: n_f^2 + 3.05863 \: n_f^3)\: a_s^4 \nonumber \\&& + (547791 - 147144 \: n_f + 11176.3 \: n_f^2 + 77.4206 \: n_f^3 - 0.636764 \: n_f^4)\: a_s^5 \nonumber \\&& + O(a_s^6) \\
    \gamma_{2,1}^{\mathcal{D}} &=& -2\: a_s + (-22.5556 + 1.96296 \: n_f)\: a_s^2 \nonumber \\&& + (-385.466 + 66.1992 \: n_f + 0.532922 \: n_f^2)\: a_s^3 \nonumber \\&& + (-6437.94 + 1751.72 \: n_f - 71.158 \: n_f^2 - 1.1502 \: n_f^3)\: a_s^4 \nonumber \\&& + (-147044 + 43307.7 \: n_f - 3728.02 \: n_f^2 - 15.8471 \: n_f^3 + 0.293057 \: n_f^4)\: a_s^5 \nonumber \\&& + O(a_s^6) \\
    \gamma_{2,0}^{\mathcal{D}} &=& -0.666667\: a_s + (-9.50617 + 0.481481 \: n_f)\: a_s^2 \nonumber \\&& + (-170.654 + 24.8232 \: n_f + 0.3107 \: n_f^2)\: a_s^3 + O(a_s^4) \\
    \gamma_{1,1} &=& 3.55556\: a_s + (48.3292 - 3.16049 \: n_f)\: a_s^2 \nonumber \\&& + (859.448 - 133.438 \: n_f - 1.22908 \: n_f^2)\: a_s^3 \nonumber \\&& + (16663.2 - 3747.55 \: n_f + 117.781 \: n_f^2 + 1.90843 \: n_f^3)\: a_s^4 \nonumber \\&& + (400747 - 103837 \: n_f + 7448.29 \: n_f^2 + 61.5735 \: n_f^3 - 0.343707 \: n_f^4)\: a_s^5 \nonumber \\&& + O(\: a_s^6) \\
    \gamma_{1,0}^{\mathcal{D}} &=& -1.77778\: a_s + (-24.1646 + 1.58025 \: n_f)\: a_s^2 \nonumber \\&& + (-429.724 + 66.7191 \: n_f + 0.61454 \: n_f^2)\: a_s^3 \nonumber \\&& + (-8331.61 + 1873.78 \: n_f - 58.8907 \: n_f^2 - 0.954217 \: n_f^3)\: a_s^4 \nonumber \\&& + (-200373 + 51918.3 \: n_f - 3724.15 \: n_f^2 - 30.7867 \: n_f^3 + 0.171853 \: n_f^4)\: a_s^5 \nonumber \\&& + O(a_s^6)
\end{eqnarray}
The values for $\gamma_{2,1}^{\mathcal{D}, (2)}$ and $\gamma_{2,0}^{\mathcal{D}, (2)}$ at order $a_s^3$ 
can be compared with those presented in \cite{Kniehl:2020nhw}, 
where the authors used numerical methods for their determination. 
We find agreement with their values of $\gamma_{2,1}^{\mathcal{D}, (2)}$ and $\gamma_{2,0}^{\mathcal{D}, (2)}$, in the latter case within the uncertainties of their numerical result~\footnote{
The numerical errors in the parentheses in  Eq.~(\ref{eq:refKV}) have been provided by the authors of~\cite{Kniehl:2020nhw} in private communication.}
\begin{equation}
\label{eq:refKV}
\gamma_{2,0}^{\mathcal{D}, (2)}\biggr|_{\text{ref.~\cite{Kniehl:2020nhw}}} \,=\,
-170.641(12)+24.822(2) \: n_f+0.3107(1) \: n_f^2
\, .
\end{equation}

%
\section{Conclusion and outlook}
\label{sec:conclusion}
\setcounter{equation}{0}

We have studied the renormalization of non-singlet quark operators  appearing in deep-inelastic scattering, including total derivative operators. 
The mixing of operators under renormalization and the explicit form of the 
anomalous dimension matrix depends on the choice of a basis for the operators involving total derivatives. 
We have discussed three choices, which have appeared in the literature, one in terms of Gegenbauer polynomials and two based on counting (total) derivatives. We have shown how to transform the anomalous dimension matrices from one basis to the other, thereby connecting as of yet unrelated results in the literature and using the results in the different bases for mutual cross-checks up to the three-loop level.
This provides highly non-trivial tests of the respective computations, 
which have applied entirely different techniques, for one exploiting conformal symmetry and a dedicated determination of the conformal anomaly, 
or, otherwise, evaluating Feynman diagrams of the respective OMEs directly.
The result for the three-loop evolution kernel for flavor-non-singlet operators in off-forward kinematics is thus established by independent methods. 
The quoted analytic expressions in moment space in the Gegenbauer and the total derivative basis in terms of harmonic sums are new.

Moreover, we have exploited consistency relations between the off-forward anomalous dimensions in the chiral limit 
to derive new results at the four- and five-loop level for the anomalous dimension matrices, 
presenting the leading-$\nf$ terms of the mixing matrix of non-singlet quark operators 
as well as the full QCD result for some low-$N$ fixed moments in the \MSb\ renormalization scheme.
With an additional scheme transformation to the RI scheme, these 
will also become useful in studies of the hadron structure with lattice QCD approaches.

The algorithms presented in this paper, i.e. consistency relations combined 
with direct Feynman diagram computations of the respective OMEs, 
are expected to work also beyond the leading-$\nf$ or the planar limit discussed here.
They allow for an automation of the calculations with the help of various computer algebra programs, such as {\sc Forcer} under {\sc Form} for the 
computation of massless two-point functions up to four loops or the symbolic summation with the help of {\sc Sigma}.
Furthermore, the approach can be adapted to the calculation of mixing matrices between other types of operator in QCD, like flavor-singlet operators, or to different models all-together, like gradient operators in scalar field theories. 
We leave these aspects for future studies.

%
\subsection*{Acknowledgements}
We thank V.~Braun, J.~Gracey and A.~Manashov for useful discussions and for comments on the manuscript. 
The Feynman diagrams have been drawn with the packages 
{\sc Axodraw} \cite{Vermaseren:1994je} and {\sc Jaxo\-draw} \cite{Binosi:2003yf}.

This work has been supported by Deutsche Forschungsgemeinschaft (DFG) through the Research Unit FOR 2926, ``Next Generation pQCD for
Hadron Structure: Preparing for the EIC'', project number 40824754 and DFG grant $\text{MO~1801/4-1}$.

\bigskip

\appendix

 ---------------------------------------------------------------------

\renewcommand{\theequation}{\ref{sec:appA}.\arabic{equation}}
\setcounter{equation}{0}
\renewcommand{\thefigure}{\ref{sec:appA}.\arabic{figure}}
\setcounter{figure}{0}
\renewcommand{\thetable}{\ref{sec:appA}.\arabic{table}}
\setcounter{table}{0}
\section{$SU(\nc)$ anomalous dimensions}
\label{sec:appA}
Here we present the off-diagonal elements of the $N=5$ mixing matrix for a general gauge group $SU(\nc)$. Starting from the 4-loop level, the following color factors contribute
\begin{eqnarray}
    \frac{d^{abcd}_Fd^{abcd}_A}{N_F} &=& \frac{(n_c^2+6)(n_c^2-1)}{48}\, , \\
    \frac{d^{abcd}_Fd^{abcd}_F}{N_F} &=& \frac{(n_c^4-6n_c^2+18)(n_c^2-1)}{96n_c^3}\, , \\
    \frac{d^{abcd}_Ad^{abcd}_A}{N_A} &=& \frac{n_c^2(n_c^2+36)}{24}\, .
\end{eqnarray}
The subscripts $F$ and $A$ denote the fundamental and adjoint representation, and $N_F=\nc$ and $N_A=\nc^2-1$ represent their dimensions. 
Hence in QCD $d^{abcd}_Fd^{abcd}_A/N_F = 5/2$,
$d^{abcd}_Fd^{abcd}_F/N_F = 5/36$ and 
$d^{abcd}_Ad^{abcd}_A/N_A = 135/8$, 
see e.g. \cite{Moch:2018wjh} for more details.

\subsection{Two-loop anomalous dimensions}

\begin{eqnarray}
    \gamma_{4,3}^{\mathcal{D}, (1)} &=& -\frac{191}{30} C_F C_A + \frac{997}{1080}C_F^2 + \frac{17}{10} n_f C_F  \, , \\[6pt]
    \gamma_{4,2}^{\mathcal{D}, (1)} &=& - \frac{7999}{1800}C_F^2+\gamma_{4,2}^{\mathcal{D}, (1)}\biggr\vert_{\rm nlc}+\frac{82}{135}n_fC_F \, , \\[6pt]
    \gamma_{4,1}^{\mathcal{D}, (1)} &=& \frac{5339}{225}C_FC_A - \frac{37769}{27000}C_F^2 + \frac{23}{90}n_fC_F - \gamma_{4,2}^{\mathcal{D}, (1)}\biggr\vert_{\rm nlc} \, , \\[6pt]
    \gamma_{4,0}^{\mathcal{D}, (1)} &=& - \frac{931}{1080}C_FC_A + \frac{14}{675}C_F^2 + \frac{43}{540}n_fC_F \, , \\[6pt]
    \gamma_{3,2}^{\mathcal{D}, (1)} &=& -\frac{1391}{225} C_F C_A + \frac{343}{375}C_F^2 + \frac{362}{225} n_f C_F  \, , \\[6pt]
    \gamma_{3,1}^{\mathcal{D}, (1)} &=& - \frac{211}{100}C_FC_A - \frac{1153}{9000}C_F^2 + \frac{13}{25}n_fC_F \, , \\[6pt]
    \gamma_{3,0}^{\mathcal{D}, (1)} &=& - \frac{958}{675}C_FC_A + \frac{772}{3375}C_F^2 + \frac{106}{675}n_fC_F \, , \\[6pt]
    \gamma_{2,1}^{\mathcal{D}, (1)} &=&  -\frac{53}{9} C_F C_A + \frac{9}{16}C_F^2 + \frac{53}{36} n_f C_F \, , \\[6pt]
    \gamma_{2,0}^{\mathcal{D}, (1)} &=& - \frac{97}{36}C_FC_A + \frac{103}{144}C_F^2 + \frac{13}{36}n_fC_F \, , \\[6pt]
    \gamma_{1,0}^{\mathcal{D}, (1)} &=& - \frac{188}{27}C_FC_A + \frac{56}{27}C_F^2 + \frac{32}{27}n_fC_F \, .
\end{eqnarray}

\subsection{Three-loop anomalous dimensions}

\begin{eqnarray}
    \gamma_{4,3}^{\mathcal{D}, (2)} &=& C_FC_A^2  \Big(  - \frac{6642011}{194400} + \frac{2}{3}\zeta_3 \Big)
       + C_F^2C_A  \Big(  - \frac{2291861}{972000} - 2\zeta_3 \Big)
     \nonumber \\&&  + C_F^3  \Big( \frac{4602059}{972000} + \frac{4}{3}\zeta_3 \Big)
        + n_fC_FC_A  \Big( \frac{28703}{12150} + \frac{40}{3}\zeta_3 \Big)
       \nonumber \\&&+ n_fC_F^2  \Big( \frac{6516293}{486000} - \frac{40}{3}\zeta_3 \Big)
       + n_f^2C_F  \Big( \frac{3131}{8100} \Big) \, , \\[6pt]
    \gamma_{4,2}^{\mathcal{D}, (2)} &=& \frac{1312}{6075}n_f^2C_F+\Tilde{\gamma}_{4,2}^{\mathcal{D}, (2)} \, , \\[6pt]
    \gamma_{4,1}^{\mathcal{D}, (2)} &=&C_FC_A^2  \Big(  - \frac{2331353}{129600} + \frac{6}{5}\zeta_3 \Big)
       + C_F^2C_A  \Big(  - \frac{1620737}{243000} - \frac{18}{5}\zeta_3 \Big)
      \nonumber \\&& + C_F^3  \Big( \frac{3600253}{540000} + \frac{12}{5}\zeta_3 \Big)
        + n_fC_FC_A  \Big( \frac{124163}{97200} + 8\zeta_3 \Big)
       \nonumber \\&&+ n_fC_F^2  \Big( \frac{3839747}{486000} - 8\zeta_3 \Big)
      + n_f^2C_F  \Big( \frac{1181}{8100} \Big) - \Tilde{\gamma}_{4,2}^{\mathcal{D}, (2)} \, , \\[6pt]
    \gamma_{4,0}^{\mathcal{D}, (2)} &=& C_FC_A^2  \Big(  - \frac{5439839}{777600} - \frac{19}{5}\zeta_3 \Big)
       + C_F^2C_A  \Big( \frac{11200081}{1944000} + \frac{57}{5}\zeta_3 \Big)
      \nonumber \\&& + C_F^3  \Big(  - \frac{14346169}{6480000} - \frac{38}{5}\zeta_3 \Big)
        + n_fC_FC_A  \Big( \frac{155383}{194400} + \frac{4}{3}\zeta_3 \Big)
       \nonumber \\&&+ n_fC_F^2  \Big( \frac{34921}{54000} - \frac{4}{3}\zeta_3 \Big)
       + n_f^2C_F  \Big( \frac{4841}{48600} \Big) \, , \\[6pt]
    \gamma_{3,2}^{\mathcal{D}, (2)} &=& C_FC_A^2  \Big(  - \frac{1555453}{40500} - \frac{32}{25}\zeta_3 \Big)
       + C_F^2C_A  \Big( \frac{755417}{67500} + \frac{96}{25}\zeta_3 \Big)
       \nonumber \\&&+ C_F^3  \Big(  - \frac{3114527}{2025000} - \frac{64}{25}\zeta_3 \Big)
        + n_fC_FC_A  \Big( \frac{63841}{20250} + \frac{64}{5}\zeta_3 \Big)
      \nonumber \\&& + n_fC_F^2  \Big( \frac{197764}{16875} - \frac{64}{5}\zeta_3 \Big)
       + n_f^2C_F  \Big( \frac{3947}{10125} \Big) \, , \\[6pt]
    \gamma_{3,1}^{\mathcal{D}, (2)} &=& C_FC_A^2  \Big(  - \frac{328189}{54000} + \frac{78}{25}\zeta_3 \Big)
       + C_F^2C_A  \Big(  - \frac{30661727}{1620000} - \frac{234}{25}\zeta_3 \Big)
      \nonumber \\&& + C_F^3  \Big( \frac{22052171}{1800000} + \frac{156}{25}\zeta_3 \Big)
       + n_fC_FC_A  \Big( \frac{2351}{3375} + \frac{24}{5}\zeta_3 \Big)
      \nonumber \\&& + n_fC_F^2  \Big( \frac{9010739}{1620000} - \frac{24}{5}\zeta_3 \Big)
       + n_f^2C_F  \Big( \frac{734}{3375} \Big) \, , \\[6pt]
    \gamma_{3,0}^{\mathcal{D}, (2)} &=& C_FC_A^2  \Big(  - \frac{319316}{30375} - \frac{376}{75}\zeta_3 \Big)
       + C_F^2C_A  \Big( \frac{18468637}{2430000} + \frac{376}{25}\zeta_3 \Big)
       \nonumber \\&&+ C_F^3  \Big(  - \frac{52103569}{24300000} - \frac{752}{75}\zeta_3 \Big)
        + n_fC_FC_A  \Big( \frac{81233}{60750} + \frac{32}{15}\zeta_3 \Big)
       \nonumber \\&&+ n_fC_F^2  \Big( \frac{1302883}{1215000} - \frac{32}{15}\zeta_3 \Big)
       + n_f^2C_F  \Big( \frac{4411}{30375} \Big) \, , \\[6pt]
    \gamma_{2,1}^{\mathcal{D}, (2)} &=& C_FC_A^2  \Big(  - \frac{73331}{2592} + 3\zeta_3 \Big)
       + C_F^2C_A  \Big(  - \frac{78193}{5184} - 9\zeta_3 \Big)
       \nonumber \\&&+ C_F^3  \Big( \frac{23185}{1728} + 6\zeta_3 \Big)
       + n_fC_FC_A  \Big( \frac{4067}{1296} + 12\zeta_3 \Big)
       + n_fC_F^2  \Big( \frac{31481}{2592} - 12\zeta_3 \Big)
       \nonumber \\&&+ n_f^2C_F  \Big( \frac{259}{648} \Big) \, , \\[6pt]
    \gamma_{2,0}^{\mathcal{D}, (2)} &=& C_FC_A^2  \Big(  - \frac{47225}{2592} - 7\zeta_3 \Big)
       + C_F^2C_A  \Big( \frac{56393}{5184} + 21\zeta_3 \Big)
      \nonumber \\&& + C_F^3  \Big(  - \frac{130}{81} - 14\zeta_3 \Big)
       + n_fC_FC_A  \Big( \frac{3335}{1296} + 4\zeta_3 \Big)
       + n_fC_F^2  \Big( \frac{2803}{1296} - 4\zeta_3 \Big)
      \nonumber \\&& + n_f^2C_F  \Big( \frac{151}{648} \Big) \, , \\[6pt]
       \gamma_{1,0}^{\mathcal{D}, (2)} &=& C_FC_A^2  \Big( -\frac{10460}{243} - \frac{32}{3}\zeta_3 \Big)
       + C_F^2C_A  \Big(   \frac{4264}{243} + 32\zeta_3 \Big)
       + C_F^3  \Big(   \frac{280}{243} - \frac{64}{3}\zeta_3 \Big)
       \nonumber \\&&+ n_fC_FC_A  \Big(   \frac{1564}{243} + \frac{32}{3}\zeta_3 \Big)
       + n_fC_F^2  \Big(   \frac{1706}{243} - \frac{32}{3}\zeta_3 \Big)
       + n_f^2C_F  \Big(   \frac{112}{243} \Big)
       \, .
\end{eqnarray}

\subsection{Four-loop anomalous dimensions}

\begin{eqnarray}
    \gamma_{1,0}^{\mathcal{D}, (3)} &=& 
       \frac{d^{abcd}_Fd^{abcd}_A}{N_F}  \Big( \frac{368}{9} - \frac{2560}{9}\zeta_5 - \frac{992}{9}\zeta_3 \Big)
     \nonumber \\&&  + C_FC_A^3  \Big(  - \frac{867065}{2187} + \frac{6080}{27}\zeta_5 + \frac{176}{3}\zeta_4 - \frac{17468}{81}\zeta_3 \Big)
       \nonumber \\&&+ C_F^2C_A^2  \Big( \frac{813032}{2187} - \frac{2240}{9}\zeta_5 - 176\zeta_4 + \frac{12872}{27}\zeta_3 \Big)
     \nonumber \\&&  + C_F^3C_A  \Big(  - \frac{119338}{2187} - \frac{640}{3}\zeta_5 + \frac{352}{3}\zeta_4 - \frac{15520}{81}\zeta_3 \Big)
       \nonumber \\&&+ C_F^4  \Big(  - \frac{97196}{2187} + \frac{640}{3}\zeta_5 - \frac{5440}{81}\zeta_3 \Big)
      \nonumber \\&& + n_f\frac{d^{abcd}_Fd^{abcd}_F}{N_F}  \Big(  - \frac{416}{9} + \frac{1280}{9}\zeta_5 - \frac{512}{9}\zeta_3 \Big)
       \nonumber \\&&+ n_fC_FC_A^2  \Big( \frac{26509}{243} - \frac{2240}{27}\zeta_5 - \frac{208}{3}\zeta_4 + \frac{2020}{9}\zeta_3 \Big)
       \nonumber \\&&+ n_fC_F^2C_A  \Big( \frac{88874}{2187} - \frac{160}{9}\zeta_5 + \frac{272}{3}\zeta_4 - \frac{6464}{27}\zeta_3 \Big)\nonumber \\&&
       + n_fC_F^3  \Big(  - \frac{95456}{2187} + \frac{320}{3}\zeta_5 - \frac{64}{3}\zeta_4 + \frac{1264}{81}\zeta_3 \Big)
       \nonumber \\&&+ n_f^2C_FC_A  \Big(  - \frac{3175}{729} + \frac{32}{3}\zeta_4 - \frac{64}{3}\zeta_3 \Big)\nonumber \\&&
       + n_f^2C_F^2  \Big(  - \frac{12472}{2187} - \frac{32}{3}\zeta_4 + \frac{64}{3}\zeta_3 \Big)
       + n_f^3C_F  \Big( \frac{512}{2187} - \frac{64}{81}\zeta_3 \Big) \, , \\[6pt]
    \gamma_{2,1}^{\mathcal{D}, (3)} &=&
       \frac{d^{abcd}_Fd^{abcd}_A}{N_F}  \Big(  - \frac{481}{9} + 1380\zeta_5 - 624\zeta_3 \Big)
       + C_FC_A^3  \Big(  - \frac{5725127}{23328} - 230\zeta_5 \nonumber \\&& - \frac{33}{2}\zeta_4 + \frac{17141}{72}\zeta_3 \Big)
       + C_F^2C_A^2  \Big( \frac{29201815}{186624} + 645\zeta_5 + \frac{99}{2}\zeta_4 - \frac{6967}{9}\zeta_3 \Big)
       \nonumber \\&&+ C_F^3C_A  \Big(  - \frac{1128755}{31104} - 540\zeta_5 - 33\zeta_4 + \frac{43153}{72}\zeta_3 \Big)\nonumber \\&&
       + C_F^4  \Big(  - \frac{15839393}{248832} + 240\zeta_5 - \frac{500}{3}\zeta_3 \Big)
       + n_f\frac{d^{abcd}_Fd^{abcd}_F}{N_F}  \Big( \frac{7}{9} + 160\zeta_5 - 104\zeta_3 \Big)\nonumber \\&&
       + n_fC_FC_A^2  \Big( \frac{223465}{2592} - \frac{280}{3}\zeta_5 - 63\zeta_4 + \frac{3403}{18}\zeta_3 \Big)
       + n_fC_F^2C_A  \Big( \frac{3855247}{46656} - 20\zeta_5 \nonumber \\&& + 57\zeta_4 - \frac{1897}{18}\zeta_3 \Big)
       + n_fC_F^3  \Big(  - \frac{1549993}{31104} + 120\zeta_5 + 6\zeta_4 - \frac{223}{3}\zeta_3 \Big)
      \nonumber \\&& + n_f^2C_FC_A  \Big(  - \frac{9649}{1944} + 12\zeta_4 - \frac{226}{9}\zeta_3 \Big)
       + n_f^2C_F^2  \Big(  - \frac{171745}{23328} - 12\zeta_4 + \frac{226}{9}\zeta_3 \Big)
      \nonumber \\&& + n_f^3C_F  \Big( \frac{2401}{11664} - \frac{8}{9}\zeta_3 \Big) \, , \\[6pt]
    \gamma_{3,2}^{\mathcal{D}, (3)} &=& 
       \frac{d^{abcd}_Fd^{abcd}_A}{N_F}  \Big(  - \frac{292963}{900} + \frac{21184}{15}\zeta_5 - \frac{12784}{15}\zeta_3 \Big)
    \nonumber \\&&   + C_FC_A^3  \Big(  - \frac{1009689359}{7290000} - \frac{5552}{45}\zeta_5 + \frac{176}{25}\zeta_4  + \frac{136949}{3375}
         \zeta_3 \Big)
       \nonumber \\&&+ C_F^2C_A^2  \Big(  - \frac{359381531}{24300000} + \frac{160}{3}\zeta_5 - \frac{528}{25}\zeta_4 - \frac{88256}{1875}
         \zeta_3 \Big)
       \nonumber \\&&+ C_F^3C_A  \Big(  - \frac{522421339}{7290000} + \frac{5248}{15}\zeta_5 + \frac{352}{25}\zeta_4 - \frac{151912}{1125}
         \zeta_3 \Big)
       \nonumber \\&&+ C_F^4  \Big( \frac{9662914729}{121500000} - \frac{2432}{15}\zeta_5 + \frac{366368}{16875}\zeta_3 \Big) \nonumber \\&&
       + n_f\frac{d^{abcd}_Fd^{abcd}_F}{N_F}  \Big(  - \frac{4193}{90} + \frac{1152}{5}\zeta_5 - \frac{9328}{75}\zeta_3 \Big)
      \nonumber \\&& + n_fC_FC_A^2  \Big( \frac{19831099}{202500} - \frac{1456}{15}\zeta_5 - \frac{1792}{25}\zeta_4  + \frac{235459}{1125}
         \zeta_3 \Big)
       \nonumber \\&&+ n_fC_F^2C_A  \Big( \frac{367343971}{6075000} - \frac{64}{3}\zeta_5 + \frac{1856}{25}\zeta_4 - \frac{17924}{125}
         \zeta_3 \Big)\nonumber \\&&
       + n_fC_F^3  \Big(  - \frac{87346403}{1822500} + 128\zeta_5 - \frac{64}{25}\zeta_4 - \frac{21352}{375}\zeta_3 \Big)
       \nonumber \\&& + n_f^2C_FC_A  \Big(  - \frac{2581429}{607500} + \frac{64}{5}\zeta_4 - \frac{2032}{75}\zeta_3 \Big)\nonumber \\&&
       + n_f^2C_F^2  \Big(  - \frac{6850267}{759375} - \frac{64}{5}\zeta_4 + \frac{2032}{75}\zeta_3 \Big)
       + n_f^3C_F  \Big( \frac{191989}{911250} - \frac{128}{135}\zeta_3 \Big) \, , \\[6pt]
    \gamma_{4,3}^{\mathcal{D}, (3)} &=& \frac{d^{abcd}_Fd^{abcd}_A}{N_F}  \Big(  - \frac{3138061}{5400} + \frac{22840}{9}\zeta_5 - \frac{315746}{225}\zeta_3 \Big)
      \nonumber \\&& + C_FC_A^3  \Big(  - \frac{339793967}{34992000} - \frac{12896}{27}\zeta_5 - \frac{11}{3}\zeta_4  + \frac{451817}{2025}
         \zeta_3 \Big)
       \nonumber \\&&+ C_F^2C_A^2  \Big(  - \frac{69304250621}{174960000} + \frac{9506}{9}\zeta_5 + 11\zeta_4 - \frac{348659}{675}
         \zeta_3 \Big)\nonumber \\&&
       + C_F^3C_A  \Big( \frac{170996951743}{437400000} - \frac{6488}{9}\zeta_5 - \frac{22}{3}\zeta_4 + \frac{62749}{300}
         \zeta_3 \Big)
       \nonumber \\&&+ C_F^4  \Big(  - \frac{652403846867}{3499200000} + \frac{3184}{9}\zeta_5 - \frac{194876}{2025}\zeta_3 \Big)
      \nonumber \\&& + n_f\frac{d^{abcd}_Fd^{abcd}_F}{N_F}  \Big(  - \frac{57823}{675} + \frac{832}{3}\zeta_5 - \frac{29552}{225}\zeta_3 \Big)
      \nonumber \\&& + n_fC_FC_A^2  \Big( \frac{170258047}{1749600} - \frac{896}{9}\zeta_5  - \frac{218}{3}\zeta_4 + \frac{9409}{45}\zeta_3 \Big)
     \nonumber \\&&  + n_fC_F^2C_A  \Big( \frac{2705226287}{43740000} - \frac{200}{9}\zeta_5 + \frac{214}{3}\zeta_4 - \frac{85819}{675}
         \zeta_3 \Big)\nonumber \\&&
       + n_fC_F^3  \Big(  - \frac{17459684723}{437400000} + \frac{400}{3}\zeta_5 + \frac{4}{3}\zeta_4 - \frac{48094}{675}
         \zeta_3 \Big)
      \nonumber \\&& + n_f^2C_FC_A  \Big(  - \frac{580649}{145800} + \frac{40}{3}\zeta_4 - \frac{3836}{135}\zeta_3 \Big)\nonumber \\&&
       + n_f^2C_F^2  \Big(  - \frac{210054821}{21870000} - \frac{40}{3}\zeta_4 + \frac{3836}{135}\zeta_3 \Big)
       + n_f^3C_F  \Big( \frac{154397}{729000} - \frac{80}{81}\zeta_3 \Big)\, .
\end{eqnarray}

\subsection{Five-loop anomalous dimensions}

\begin{eqnarray}
    \gamma_{1,0}^{\mathcal{D}, (4)} &=&
       C_A\frac{d^{abcd}_Fd^{abcd}_A}{N_F}  \Big( \frac{41384}{81} + 6384\zeta_7 + \frac{70400}{27}\zeta_6 - \frac{646480}{81}\zeta_5 + \frac{5456}{9}
         \zeta_4 \nonumber \\&& - \frac{277760}{81}\zeta_3 + \frac{42176}{27}\zeta_3^2 \Big)
       + C_F\frac{d^{abcd}_Ad^{abcd}_A}{N_A}  \Big( \frac{7672}{81} - \frac{9520}{9}\zeta_7 + \frac{29200}{27}\zeta_5 \nonumber \\&& - \frac{352}{3}\zeta_4 - \frac{6032}{27}\zeta_3
          - \frac{3008}{3}\zeta_3^2 \Big)
       + C_F\frac{d^{abcd}_Fd^{abcd}_A}{N_F}  \Big(  - \frac{11984}{81} - \frac{38528}{9}\zeta_7 \nonumber \\&& - \frac{88160}{81}\zeta_5 + \frac{366752}{81}\zeta_3 - 
         \frac{3200}{3}\zeta_3^2 \Big)
       + C_FC_A^4  \Big(  - \frac{266532611}{78732} - \frac{89488}{27}\zeta_7 \nonumber \\&& - \frac{167200}{81}\zeta_6 + \frac{1551104}{243}
         \zeta_5 + \frac{110960}{81}\zeta_4 - \frac{1294072}{729}\zeta_3 - \frac{37456}{81}\zeta_3^2 \Big)\nonumber \\&&
       + C_F^2C_A^3  \Big( \frac{110112362}{19683} + \frac{165928}{27}\zeta_7 + \frac{61600}{27}\zeta_6 - \frac{1820312}{243}
         \zeta_5 - \frac{85484}{27}\zeta_4 \nonumber \\&& + \frac{2057768}{729}\zeta_3 + \frac{35200}{27}\zeta_3^2 \Big)
       + C_F^3C_A^2  \Big(  - \frac{31670203}{6561} - 7840\zeta_7 + \frac{17600}{9}\zeta_6 \nonumber \\&& - \frac{30848}{27}\zeta_5
          + \frac{114736}{81}\zeta_4 + \frac{501596}{243}\zeta_3 - \frac{15488}{9}\zeta_3^2 \Big)\nonumber \\&&
       + C_F^4C_A  \Big( \frac{40931372}{19683} + 9968\zeta_7 - \frac{17600}{9}\zeta_6 - \frac{71120}{27}\zeta_5 + 
         \frac{29920}{81}\zeta_4 \nonumber \\&& - \frac{800296}{243}\zeta_3 + 1536\zeta_3^2 \Big)
       + C_F^5  \Big(  - \frac{4653188}{19683} - 4256\zeta_7 + \frac{278720}{81}\zeta_5 \nonumber \\&& + \frac{401392}{729}\zeta_3 - 
         \frac{6272}{9}\zeta_3^2 \Big)
       + n_f\frac{d^{abcd}_Fd^{abcd}_A}{N_F}  \Big(  - \frac{11048}{27} + 1232\zeta_7 - \frac{12800}{27}\zeta_6 \nonumber \\&& + \frac{108640}{81}\zeta_5 + \frac{256}{9}
         \zeta_4 - \frac{21856}{81}\zeta_3 + \frac{12544}{27}\zeta_3^2 \Big)\nonumber \\&&
       + n_fC_A\frac{d^{abcd}_Fd^{abcd}_F}{N_F}  \Big(  - \frac{103912}{81} + \frac{14560}{9}\zeta_7 - \frac{35200}{27}\zeta_6 + \frac{261440}{81}\zeta_5
         \nonumber \\&& + \frac{2816}{9}\zeta_4 - \frac{125696}{81}\zeta_3 - \frac{7936}{27}\zeta_3^2 \Big)
       + n_fC_F\frac{d^{abcd}_Fd^{abcd}_F}{N_F}  \Big( \frac{85376}{81} - \frac{17920}{9}\zeta_7 \nonumber \\&& + \frac{325120}{81}\zeta_5 - \frac{164416}{81}\zeta_3 - \frac{4096}{9}\zeta_3^2 \Big)
       + n_fC_FC_A^3  \Big( \frac{24423290}{19683} + \frac{19544}{27}\zeta_7 \nonumber \\&& + \frac{92000}{81}\zeta_6 - \frac{694540}{243}
         \zeta_5 - \frac{137384}{81}\zeta_4 + \frac{2157154}{729}\zeta_3 + \frac{13904}{81}\zeta_3^2 \Big)\nonumber \\&&
       + n_fC_F^2C_A^2  \Big(  - \frac{15291499}{26244} - \frac{5600}{27}\zeta_7 - \frac{6800}{27}\zeta_6 + \frac{126272}{243}
         \zeta_5 + \frac{57268}{27}\zeta_4 \nonumber \\&& - \frac{780800}{243}\zeta_3 - \frac{12448}{27}\zeta_3^2 \Big)
       + n_fC_F^3C_A  \Big( \frac{1687541}{6561} + \frac{2240}{3}\zeta_7 - \frac{4000}{3}\zeta_6 \nonumber \\&& + \frac{229016}{81}\zeta_5 - 
         \frac{24128}{81}\zeta_4 - \frac{210034}{243}\zeta_3 + \frac{1984}{3}\zeta_3^2 \Big)\nonumber \\&&
       + n_fC_F^4  \Big(  - \frac{912482}{19683} - \frac{4480}{3}\zeta_7 + \frac{3200}{9}\zeta_6 + \frac{8240}{81}\zeta_5 - 
         \frac{10624}{81}\zeta_4 \nonumber \\&& + \frac{231760}{243}\zeta_3 - \frac{3328}{9}\zeta_3^2 \Big)
       + n_f^2\frac{d^{abcd}_Fd^{abcd}_F}{N_F}  \Big( \frac{21872}{81} + \frac{6400}{27}\zeta_6 - \frac{26240}{81}\zeta_5 \nonumber \\&& - \frac{896}{9}\zeta_4 - \frac{17824}{81}
         \zeta_3 + \frac{1024}{27}\zeta_3^2 \Big)
       + n_f^2C_FC_A^2  \Big(  - \frac{315700}{6561} - \frac{11200}{81}\zeta_6 \nonumber \\&& + \frac{26672}{243}\zeta_5 + 392\zeta_4
          - \frac{107134}{243}\zeta_3 - \frac{12736}{81}\zeta_3^2 \Big)\nonumber \\&&
       + n_f^2C_F^2C_A  \Big(  - \frac{166127}{2187} - \frac{800}{27}\zeta_6 + \frac{14272}{81}\zeta_5 - \frac{10376}{27}
         \zeta_4 + \frac{42508}{243}\zeta_3 \nonumber \\&& + \frac{6976}{27}\zeta_3^2 \Big)
       + n_f^2C_F^3  \Big(  - \frac{1082297}{13122} + \frac{1600}{9}\zeta_6 - \frac{27776}{81}\zeta_5 - \frac{536}{81}\zeta_4
         \nonumber \\&& + \frac{72896}{243}\zeta_3 - \frac{896}{9}\zeta_3^2 \Big)
       + n_f^3C_FC_A  \Big(  - \frac{168677}{39366} + \frac{2048}{81}\zeta_5 - \frac{1376}{81}\zeta_4 \nonumber \\&& - \frac{5936}{729}
         \zeta_3 \Big)
       + n_f^3C_F^2  \Big(  - \frac{132755}{19683} - \frac{256}{27}\zeta_5 + \frac{64}{3}\zeta_4 - \frac{5936}{729}\zeta_3 \Big)\nonumber \\&&
       + n_f^4C_F  \Big( \frac{2752}{19683} - \frac{64}{81}\zeta_4 + \frac{512}{729}\zeta_3 \Big) \, , \\[6pt]
       \gamma_{2,1}^{\mathcal{D}, (4)} &=& C_A\frac{d^{abcd}_Fd^{abcd}_A}{N_F}  \Big(  - \frac{663149}{162} - \frac{43547}{4}\zeta_7 - 12650\zeta_6 + \frac{2347355}{54}\zeta_5 \nonumber \\&& + 
         3432\zeta_4 - \frac{235195}{54}\zeta_3 - 10416\zeta_3^2 \Big)
      \nonumber \\&& + C_F\frac{d^{abcd}_Ad^{abcd}_A}{N_A}  \Big( \frac{51037}{162}  + \frac{3115}{4}\zeta_7 + \frac{39275}{54}\zeta_5  - 132\zeta_4 \nonumber \\&& - \frac{76387}{54}
         \zeta_3 - 328\zeta_3^2 \Big)
       + C_F\frac{d^{abcd}_Fd^{abcd}_A}{N_F}  \Big( \frac{2180047}{324} - 13706\zeta_7 \nonumber \\&& - \frac{31785}{2}\zeta_5  + \frac{1139143}{108}\zeta_3 + 
         9870\zeta_3^2 \Big)
     \nonumber \\&&  + C_FC_A^4  \Big(  - \frac{2566577183}{3359232} - \frac{2296}{3}\zeta_7 + \frac{6325}{3}\zeta_6  - \frac{5133413}{648}
         \zeta_5 \nonumber \\&& - \frac{195499}{144}\zeta_4 + \frac{36838057}{7776}\zeta_3 + \frac{6356}{3}\zeta_3^2 \Big)
     \nonumber \\&&  + C_F^2C_A^3  \Big(  - \frac{19615708519}{6718464} + \frac{88879}{12}\zeta_7 - \frac{11825}{2}\zeta_6 + 
         \frac{9916061}{432}\zeta_5 \nonumber \\&& + \frac{79391}{18}\zeta_4  - \frac{5873293}{324}\zeta_3 - 7570\zeta_3^2 \Big)
      \nonumber \\&& + C_F^3C_A^2  \Big( \frac{1361909845}{373248} - \frac{77175}{4}\zeta_7 + 4950\zeta_6 - \frac{1478507}{216}
         \zeta_5 \nonumber \\&& - \frac{489371}{144}\zeta_4 + \frac{106509385}{7776}\zeta_3 + 10534\zeta_3^2 \Big)
     \nonumber \\&&  + C_F^4C_A  \Big(  - \frac{16427221735}{26873856} + \frac{75411}{4}\zeta_7 - 2200\zeta_6 - \frac{1899115}{108}
         \zeta_5 \nonumber \\&& + \frac{2750}{3}\zeta_4 + \frac{9100805}{2592}\zeta_3 - 7562\zeta_3^2 \Big)\nonumber \\&&
       + C_F^5  \Big(  - \frac{178411603}{331776} - \frac{17661}{2}\zeta_7 + \frac{84485}{6}\zeta_5 - \frac{4632269}{864}
         \zeta_3 + 2466\zeta_3^2 \Big)\nonumber \\&&
       + n_f\frac{d^{abcd}_Fd^{abcd}_A}{N_F}  \Big(  - \frac{165313}{162} - 1554\zeta_7 + 2300\zeta_6 - \frac{3500}{3}\zeta_5 \nonumber \\&& - 468\zeta_4 - 
         \frac{12362}{9}\zeta_3 + \frac{4168}{3}\zeta_3^2 \Big)\nonumber \\&&
       + n_fC_A\frac{d^{abcd}_Fd^{abcd}_F}{N_F}  \Big( \frac{173813}{162} + 4760\zeta_7 - \frac{4400}{3}\zeta_6 - \frac{68660}{27}\zeta_5 \nonumber \\&& + 572
         \zeta_4 - \frac{27365}{27}\zeta_3 - 544\zeta_3^2 \Big)\nonumber \\&&
       + n_fC_F\frac{d^{abcd}_Fd^{abcd}_F}{N_F}  \Big(  - \frac{97597}{81} - 11060\zeta_7 + \frac{108040}{9}\zeta_5 - \frac{1726}{27}\zeta_3 - 32
         \zeta_3^2 \Big)\nonumber \\&&
       + n_fC_FC_A^3  \Big( \frac{837883325}{839808} + \frac{10507}{12}\zeta_7 + \frac{4250}{9}\zeta_6 - \frac{139930}{81}
         \zeta_5 \nonumber \\&& - \frac{72251}{72}\zeta_4 + \frac{3727589}{1944}\zeta_3 - \frac{2957}{9}\zeta_3^2 \Big)\nonumber \\&&
       + n_fC_F^2C_A^2  \Big(  - \frac{191432243}{373248} - \frac{595}{12}\zeta_7 + \frac{3775}{3}\zeta_6 - \frac{84596}{27}
         \zeta_5 \nonumber \\&& - \frac{215}{36}\zeta_4 + \frac{15259}{18}\zeta_3 + 1108\zeta_3^2 \Big)\nonumber \\&&
       + n_fC_F^3C_A  \Big( \frac{669771445}{1119744} + 840\zeta_7 - 2000\zeta_6 + \frac{259217}{54}\zeta_5 \nonumber \\&& + 
         \frac{72013}{72}\zeta_4 - \frac{14222911}{3888}\zeta_3 - 896\zeta_3^2 \Big)\nonumber \\&&
       + n_fC_F^4  \Big( \frac{653668241}{13436928} - 1680\zeta_7 + 400\zeta_6 + \frac{4615}{27}\zeta_5 - \frac{446}{3}
         \zeta_4 \nonumber \\&& + \frac{135293}{162}\zeta_3 + 184\zeta_3^2 \Big)
       + n_f^2\frac{d^{abcd}_Fd^{abcd}_F}{N_F}  \Big( \frac{14561}{81} + \frac{800}{3}\zeta_6 - \frac{5840}{27}\zeta_5 \nonumber \\&& - 152\zeta_4 - \frac{5906}{27}\zeta_3
          + \frac{128}{3}\zeta_3^2 \Big)\nonumber \\&&
       + n_f^2C_FC_A^2  \Big(  - \frac{7112759}{139968} - \frac{1400}{9}\zeta_6 + \frac{11882}{81}\zeta_5 + \frac{1151}{3}
         \zeta_4 \nonumber \\&& - \frac{86963}{216}\zeta_3 - \frac{1592}{9}\zeta_3^2 \Big)
       + n_f^2C_F^2C_A  \Big(  - \frac{582217}{17496} - \frac{100}{3}\zeta_6 + \frac{4772}{27}\zeta_5 \nonumber \\&& - \frac{549}{2}\zeta_4
          - \frac{535}{12}\zeta_3 + \frac{872}{3}\zeta_3^2 \Big)
       + n_f^2C_F^3  \Big(  - \frac{78593429}{559872} + 200\zeta_6 \nonumber \\&& - \frac{1144}{3}\zeta_5 - \frac{298}{3}\zeta_4 + 
         \frac{222397}{486}\zeta_3 - 112\zeta_3^2 \Big)\nonumber \\&&
       + n_f^3C_FC_A  \Big(  - \frac{977951}{419904} + \frac{256}{9}\zeta_5 - \frac{182}{9}\zeta_4 - \frac{1025}{243}\zeta_3 \Big)\nonumber \\&&
       + n_f^3C_F^2  \Big(  - \frac{3921833}{419904} - \frac{32}{3}\zeta_5 + \frac{226}{9}\zeta_4 - \frac{121}{9}\zeta_3 \Big)\nonumber \\&&
       + n_f^4C_F  \Big( \frac{27955}{209952} - \frac{8}{9}\zeta_4 + \frac{212}{243}\zeta_3 \Big)\, .
\end{eqnarray}

\bigskip
{\footnotesize

\bibliographystyle{JHEP}
\bibliography{omebib}

\if{1=0}
\providecommand{\href}[2]{#2}\begingroup\raggedright\endgroup
\fi
}

\end{document}